\documentclass[letterpaper,twocolumn,prl,aps,superscriptaddress,amsmath,amssymb,floatfix]{revtex4-1}
\usepackage{natbib}
\usepackage{mathptmx}
\usepackage[latin9]{inputenc}
\usepackage{color}
\usepackage{verbatim}
\usepackage{float}
\usepackage{amssymb,amsmath,amssymb,amsthm,mathrsfs}
\usepackage{graphicx}
\usepackage{esint}
\usepackage{siunitx}
\usepackage{soul}
\usepackage[unicode=true,
 bookmarks=true,bookmarksnumbered=false,bookmarksopen=false,
 breaklinks=false,pdfborder={0 0 1},backref=false,colorlinks=true]
 {hyperref}
\hypersetup{
 linkcolor=magenta,urlcolor=blue,citecolor=blue,pdfstartview={FitH},hyperfootnotes=false}

\makeatletter

\usepackage{textcomp}
\usepackage{epstopdf}

\pdfpageheight\paperheight
\pdfpagewidth\paperwidth

\@ifundefined{textcolor}{}{%
 \definecolor{BLACK}{gray}{0}
 \definecolor{WHITE}{gray}{1}
 \definecolor{RED}{rgb}{1,0,0}
 \definecolor{GREEN}{rgb}{0,1,0}
 \definecolor{BLUE}{rgb}{0,0,1}
 \definecolor{CYAN}{cmyk}{1,0,0,0}
 \definecolor{MAGENTA}{cmyk}{0,1,0,0}
 \definecolor{YELLOW}{cmyk}{0,0,1,0}
}

\usepackage{xcolor}
\usepackage{soul}
\newcommand{\bra}[1]{\ensuremath{\left\langle#1\right|}}
\newcommand{\ket}[1]{\ensuremath{\left|#1\right\rangle}}

\definecolor{blue}{rgb}{0,0,1}
\definecolor{red}{rgb}{1,0,0}
\definecolor{green}{rgb}{0,1,0}

\usepackage{soul}

\makeatother

\begin{document}
\title{Robust and optimal control of open quantum systems}

\author{Zi-Jie Chen}
\thanks{These authors contribute equally to this work.}
\affiliation{CAS Key Laboratory of Quantum Information, University of Science
and Technology of China, Hefei, Anhui 230026, P. R. China}

\author{Hongwei Huang}
\thanks{These authors contribute equally to this work.}
\affiliation{Center for Quantum Information, Institute for Interdisciplinary Information
Sciences, Tsinghua University, Beijing 100084, China}

\author{Lida Sun}
\thanks{These authors contribute equally to this work.}
\affiliation{Center for Quantum Information, Institute for Interdisciplinary Information
Sciences, Tsinghua University, Beijing 100084, China}

\author{Qing-Xuan Jie}
\affiliation{CAS Key Laboratory of Quantum Information, University of Science
and Technology of China, Hefei, Anhui 230026, P. R. China}
\affiliation{CAS Center For Excellence in Quantum Information and Quantum Physics,
University of Science and Technology of China, Hefei, Anhui 230026, China}

\author{Jie Zhou}
\affiliation{Center for Quantum Information, Institute for Interdisciplinary Information
Sciences, Tsinghua University, Beijing 100084, China}

\author{Ziyue Hua}
\affiliation{Center for Quantum Information, Institute for Interdisciplinary Information
Sciences, Tsinghua University, Beijing 100084, China}

\author{Yifang Xu}
\affiliation{Center for Quantum Information, Institute for Interdisciplinary Information
Sciences, Tsinghua University, Beijing 100084, China}

\author{Weiting Wang}
\affiliation{Center for Quantum Information, Institute for Interdisciplinary Information
Sciences, Tsinghua University, Beijing 100084, China}

\author{Guang-Can Guo}
\affiliation{CAS Key Laboratory of Quantum Information, University of Science
and Technology of China, Hefei, Anhui 230026, P. R. China}
\affiliation{CAS Center For Excellence in Quantum Information and Quantum Physics,
University of Science and Technology of China, Hefei, Anhui 230026, China}
\affiliation{Hefei National Laboratory, Hefei 230088, China.}

\author{Chang-Ling Zou}
\email{clzou321@ustc.edu.cn }
\affiliation{CAS Key Laboratory of Quantum Information, University of Science
and Technology of China, Hefei, Anhui 230026, P. R. China}
\affiliation{CAS Center For Excellence in Quantum Information and Quantum Physics,
University of Science and Technology of China, Hefei, Anhui 230026, China}
\affiliation{Hefei National Laboratory, Hefei 230088, China.}

\author{Luyan Sun}
\email{luyansun@tsinghua.edu.cn }
\affiliation{Center for Quantum Information, Institute for Interdisciplinary Information
Sciences, Tsinghua University, Beijing 100084, China}
\affiliation{Hefei National Laboratory, Hefei 230088, China.}

\author{Xu-Bo Zou}
\email{xbz@ustc.edu.cn}
\affiliation{CAS Key Laboratory of Quantum Information, University of Science
and Technology of China, Hefei, Anhui 230026, P. R. China}
\affiliation{CAS Center For Excellence in Quantum Information and Quantum Physics,
University of Science and Technology of China, Hefei, Anhui 230026, China}
\affiliation{Hefei National Laboratory, Hefei 230088, China.}

\begin{abstract}
Recent advancements in quantum technologies have highlighted the importance of mitigating system imperfections, including parameter uncertainties and decoherence effects, to improve the performance of experimental platforms. However, most of the previous efforts in quantum control are devoted to the realization of arbitrary unitary operations in a closed quantum system. Here, we improve the algorithm that suppresses system imperfections and noises, providing notably enhanced scalability for robust and optimal control of open quantum systems. Through experimental validation in a superconducting quantum circuit, we demonstrate that our approach outperforms its conventional counterpart for closed quantum systems with an ultra-low infidelity of about $0.60\%$, while the complexity of this algorithm exhibits the same scaling, with only a modest increase in the prefactor. This work represents a notable advancement in quantum optimal control techniques, paving the way for realizing quantum-enhanced technologies in practical applications.
\end{abstract}
\maketitle
\noindent \textbf{\large{}INTRODUCTION}{\large\par}

\noindent

Quantum technologies are increasingly pivotal across diverse domains, including quantum computation, quantum communication, and quantum precision measurement~\cite{preskill2018quantum}. The critical technique driving these applications is the coherent manipulation of quantum states within a quantum system. For example, the implementation of quantum gates forms the foundation of quantum computation tasks~\cite{nielsen2010quantum}, and unitary operations are essential for preparing exotic quantum states that enable quantum-enhanced sensors~\cite{degen2017quantum}. In a given quantum system, unitary operations are realized through specific control Hamiltonians by adjusting control parameters like amplitude, detuning, and pulse shape of external control fields, such as microwave and optical driving fields~\cite{d2021introduction}. Therefore, optimizing control fields to achieve a desired target unitary operation is crucial for advancing quantum technologies. Over the past decades, various quantum optimal control methods, including the shortcut to adiabaticity~\cite{Guery-Odelin2019}, composite pulse sequences~\cite{wang2012composite}, chopped random basis~\cite{caneva2011chopped}, derivative removal by adiabatic gate~\cite{motzoi2009DRAG}, and gradient ascent pulse engineering (GRAPE)~\cite{khaneja2005optimal},  have been proposed and applied in various platforms ranging from superconducting circuits~\cite{heeres2017implementing,hu2019quantum}, ion traps~\cite{nebendahl2009optimal}, neutral atoms~\cite{larrouy2020fast}, to defects in diamonds~\cite{chou2015optimal}.  More recently, machine learning approaches, such as reinforcement learning~\cite{sivak2022model} and generative adversarial algorithms~\cite{ge2020robust}, have been integrated into quantum control to find solutions for target unitary operations.

Unfortunately, the realization of high-fidelity quantum operation is limited in practical experimental systems, even when suitable optimal control pulse shapes have been solved through numerical algorithms~\cite{kim2021hardware}. One notable limitation stems from the discrepancies of Hamiltonian between the practical system and its model, including the uncertainty of parameters and the additional stray interaction between the system and the environmental degrees of freedom. Coherent errors arise when the parameter varies, due to system mis-calibration~\cite{sheldon2016characterizing}, sample aging~\cite{yan2020engineering}, and slow drift of experimental setup~\cite{proctor2020detecting}. Addressing these errors requires improved calibration precision, enhanced system stability, and the utilization of robust quantum control techniques. Apart from coherent errors, practical experimental platforms inevitably couple with the external reservoirs \cite{catelani2012decoherence}, which leads to the decoherence of quantum systems and has been recognized as another major challenge in realizing reliable quantum information processors. Therefore, suppressing these decoherence errors requires extending system lifetimes and refining optimal control algorithms tailored for open quantum systems. 

Despite the critical importance of error suppression in quantum applications, existing efforts have primarily focused on the robust control of quantum systems~\cite{wu2019learning,koswara2021quantum,skinner2003application,kobzar2012exploring}  or have relied on the time-consuming  trajectory sampling~\cite{abdelhafez2019gradient,eickbusch2022fast}, precise evolution solving~\cite{khaneja2005optimal,boutin2017resonator,schulte2011optimal,machnes2011comparing}, and machine learning~\cite{wu2019learning} techniques for open system control.  
A numerically efficient algorithm that directly considers both the parameter uncertainties and the decoherence errors in optimizing open quantum system control remains elusive due to computational challenges. 
Take the GRAPE algorithm as an example, the algorithm that solely considers the intrinsic evolution of closed systems can be described by the Schr\"{o}dinger equation. For simplicity, we refer to this as the Closed-GRAPE algorithm. This algorithm allows the computation of the objective function and gradients by calculating the evolution of a $d$-dimensional pure state represented by a $d \times 1$-dimensional column vector, where the computational complexity is $O(d^2)$. In contrast, previous GRAPE algorithms for an open system~\cite{khaneja2005optimal,boutin2017resonator,schulte2011optimal,machnes2011comparing}  describe the evolution using the Lindblad master equation. We refer to these as the Open-GRAPE algorithms, drawing on the concept and terminology from Ref.~\cite{schulte2011optimal}. These algorithms require the computation of the evolution of the density matrix~\cite{gordon2008optimal}. This involves the matrix calculation of $d\times d$-dimensional matrices, where the computational complexity is at least $O(d^3)$ or even $O(d^6)$. For multi-qubit systems, the optimization can also be implemented by solving the Bloch equation and the computational complexity remains comparable to the situation of the Lindblad equation~\cite{gershenzon2007optimal}.

Here, a numerically efficient optimal  control algorithm for open quantum systems is introduced and experimentally verified in a superconducting quantum circuit. This approach, developed based on the conventional GRAPE algorithms, can efficiently calculate the control pulses while accounting for both system parameter uncertainties and decoherence. Referred to as the approximate Open-GRAPE algorithm, our approach demonstrates superior performance over the Closed-GRAPE algorithm. In the examples we present,  the approximate Open-GRAPE algorithm exhibits an enhancement in yield, which is the probability of generating high-quality pulses from random initial pulses. In particular, when controlling a bosonic mode with an ancillary transmon qubit, our approach provides pulses with ultra-low infidelities below $0.60\%$ after just 30 trials of the approximate Open-GRAPE algorithm, surpassing the best infidelity of $1.44\%$ obtained by the Closed-GRAPE algorithm. Both numerical and experimental results highlight the advantages of our approach in achieving the best available fidelities for controlling quantum systems in practice and reducing the number of trials in optimization. Regarding computational complexity, the approximate Open-GRAPE algorithm exhibits a linear increase in computation time with the number of uncertain parameters and noise sources considered, showcasing the potential for robust control of open quantum systems with dimensions up to $10^6$ using personal computers. 

\begin{figure}
    \begin{centering}
        \includegraphics[width=0.46\textwidth]{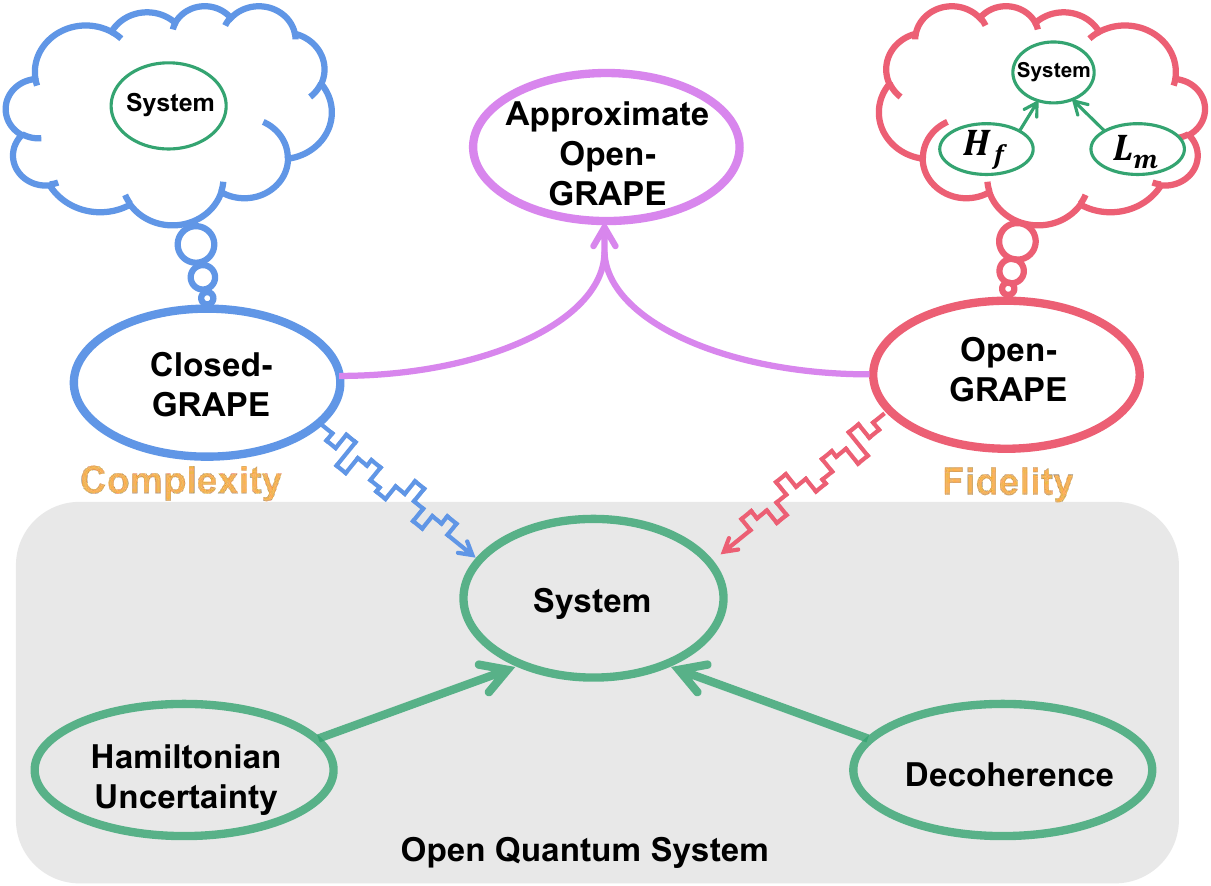}
    \end{centering}    
\caption{\label{fig:Schematic}\textbf{The schematic of the optimal control algorithms.} The shadow area represents the general model of the open quantum system considering both the uncertainty of the system Hamiltonian and the decoherence. Exhibiting low computational complexity,  Closed-GRAPE algorithm (the blue part) solely focuses on the dynamics of the ideal closed system when optimizing control pulses. 
In contrast, Open-GRAPE algorithms  (the red part) consider these two types of perturbations during the optimization process, leading to an enhancement of the operation fidelity.
The purple part corresponds to the approximate Open-GRAPE algorithm that shows both of the advantages mentioned above. 
}
\end{figure}

\smallskip{}
\smallskip{}
\noindent \textbf{\large{}RESULTS}{\large\par}

\noindent \textbf{Model}

\noindent Figure~\ref{fig:Schematic} sketches the principle of quantum control of a general open quantum system. The target system is controllable by applying external control pulses, with the total Hamiltonian of an ideal closed quantum system described by 
\begin{equation}\label{Eq1Hamiltonian}
H\left(t\right)=H_{0}+\sum_{k}u^{(k)}_{\mathrm{c}}\left(t\right)H^{(k)}_{\mathrm{c}},
\end{equation}
where $H_{0}$ is the drift Hamiltonian of the bare system and $H_{\mathrm{c}}^{(k)}$ is the control Hamiltonian with time-varying amplitude parameter $u^{(k)}_{\mathrm{c}}\left(t\right)$. For conventional quantum control algorithms, such as the GRAPE approach, the control pulse is discretized equally into $N$ steps and remains constant in each step, i.e., $u^{(k)}_{\mathrm{c}}\left(t\right)=u^{(k)}_{\mathrm{c},j}$ in the $j$-th step for $t \in [(j-1)\tau,j \tau]$, where $T$ is the duration of control pulse and $\tau=T/N$ is the step size.  Therefore, the evolution in this step is unitary with the operator $U_j=\mathrm{exp}\left[ -i \tau (H_{0}+\sum_{k}u^{(k)}_{\mathrm{c},j}H^{(k)}_{\mathrm{c}})\right]$ and the algorithm is implemented to optimize the whole evolution of this ideal system to approach the target unitary evolution as 
\begin{equation}\label{Eq2UT}
   U(T)=U_N\cdots U_2 U_1\approx U_{\mathrm{targ}}.
\end{equation}

However, there are usually imperfections in the practical quantum system. These perturbations can be divided into two categories: one is the uncertainty of the system Hamiltonian due to the parameter instability of the hardware and also the mis-calibration of the parameters; the other one is the noise due to inevitable coupling between the system and the environment. The first kind of perturbation can be described as $\sum_{m}\delta u^{(m)}_{\mathrm{f}}H^{(m)}_{\mathrm{f}}$, where $\{H_{\mathrm{f}}^{(m)}\}$ are the fluctuating Hamiltonian terms with uncertain amplitudes $\delta u_{\mathrm{f}}^{(m)}$ satisfying $\left\langle \delta u^{(m)}_{\mathrm{f}}\right\rangle =0,\left\langle (\delta u_{\mathrm{f}}^{(m)})^2\right\rangle =(\sigma_{\mathrm{f}}^{(m)})^2$. Here $\delta u_{\mathrm{f}}^{(m)}$ is a random variable that is fixed in each single-shot implementation of the system dynamics but might vary from shot to shot. $\left\langle\cdot\right\rangle$ is the average over the distribution of the random variable. Specifically, it is the ensemble average for the mis-calibration variables, while it is the temporal average for the slowly varying unstable parameters. The second kind of perturbation is usually treated as  decoherence, which can be described as Lindblad jump operators  $L_{m}$ with a noise strength $\kappa_{m}$. Considering these perturbations, as shown in the shadow area in Fig.~\ref{fig:Schematic}, the complete evolution of an open quantum system should be described by the Lindblad master equation~\cite{jacobs2014quantum}:
\begin{equation}\label{Eq3MaterEq}
    \frac{d}{dt}\rho\left( t\right)=-i \left[ H\left(t\right)+\sum_{m}\delta u^{(m)}_{\mathrm{f}}H^{(m)}_{\mathrm{f}},\rho\left( t\right) \right]+ \mathscr{L} \left( \rho\left( t\right) \right),
\end{equation}
where $\rho\left(t\right)$ is the density matrix of the system and  
\begin{equation}\label{Eq4Lindblad}
    \mathscr{L}\left(\rho\right)=\sum_{m}\kappa_{m}\left\{ L_{m}\rho L_{m}^{\dagger}-\frac{1}{2}L_{m}^{\dagger}L_{m}\rho-\frac{1}{2}\rho L_{m}^{\dagger}L_{m}\right\}
\end{equation}
is the Lindblad superoperator.

The actual performance is expected to be drastically inferior to the anticipated performance when the pulses are optimized based on the ideal closed system by Equ.~\ref{Eq1Hamiltonian} and subsequently applied to the open system by Equ.~\ref{Eq3MaterEq} in the presence of both types of perturbations. In other words, the obtained pulses are not robust against these noises, substantially deteriorating the fidelity of quantum operations. To provide a more practical evaluation of the operation quality and suppress the potential deviations from expectation, it demands an optimization algorithm that optimizes the gate while considering these perturbations, as shown in Fig.~\ref{fig:Schematic}.

\smallskip{}
\smallskip{}
\noindent \textbf{Approximate Open-GRAPE Algorithm}

\noindent  In the Closed-GRAPE algorithm, the target $U_{\mathrm{targ}}$ can be represented as a set of state transfers $\{\ket{\psi_{\mathrm{i}}^{j}}\mapsto\ket{\psi_{\mathrm{o}}^{j}}\}$, where $\ket{\psi_{\mathrm{i}}^{j}}$ and $\ket{\psi_{\mathrm{o}}^{j}}$ are the initial state and output target state respectively, and each element in the set corresponds to a constraint in the optimization of $U(T)$. In this paper, we employ the situation with a single constraint as an example for simplicity, and more general cases with multiple constraints are provided in Supplementary Materials. In Closed-GRAPE, the goal of the optimization is to maximize the objective function  as  $J_{\mathrm{close}}=|\bra{\psi_{\mathrm{o}}}U(T)\ket{\psi_{\mathrm{i}}}|^2$, which is the fidelity of the final state to the target state in the ideal closed system.

To optimize the control parameters, the estimation of the gradient $\partial J_\mathrm{close}/\partial u^{(k)}_{c,j}$ is essential. In the situation where the Closed-GRAPE applies, i.e. there is no decoherence noise or uncertain Hamiltonian parameter, the gradient can be solved efficiently and exactly. After the evolution in a closed quantum system, the final state becomes  $U_{1\rightarrow N }\rho U_{1\rightarrow N}^{\dagger}$. Here,  $U_{j \rightarrow j+i}=U_{j+i}\cdots U_{j+1}U_{j}$  ($U_0=I$ and $U_{i\rightarrow j}=I$ for $i>j$) is the unitary evolution from  $j$-th to  $j+i$-th step. Under this evolution, the gradient reads
\begin{equation}
   \frac{\partial J_\mathrm{close} }{\partial u^{(k)}_{c,j}}\propto -i\tau \bra{\psi_{\mathrm{o}}}
   U_{j+1 \rightarrow N}H^{(k)}_c\rho _{j}U_{j+1 \rightarrow N}^{\dagger}\ket{\psi_{\mathrm{o}}} +c.c., 
\end{equation}
where $\rho_{j}=\rho\left(j\tau\right)=U_{ 1 \rightarrow j} \rho_{0}U_{1\rightarrow j}^{\dagger}$ and $c.c.$ is the complex conjugate of the former part. The key idea is to calculate the gradient  through the forward propagation of the initial state (i.e., $\{ U_{1 \rightarrow j}\left|\psi_{\mathrm{i}}\right\rangle \}$) and the backward propagation of the target states (i.e., $\{U_{j+1 \rightarrow N}^{\dagger} \left|\psi_{\mathrm{o}}\right\rangle \}$), which circumvents the calculation of the differential numerically.

While the Closed-GRAPE algorithm shows an impressively fast iteration speed, its corresponding objective function and gradient exhibit inaccuracy for an open quantum system. This limits the quality of quantum operations in real experimental setups. In contrast, the Open-GRAPE algorithms~\cite{khaneja2005optimal,boutin2017resonator,schulte2011optimal,machnes2011comparing}, depicted in the red region of Fig.~\ref{fig:Schematic}, provide error-resisting quantum operations by considering the complete dynamics, thereby improving the operation fidelity. However, this also comes with higher computational complexity.

For a practical evaluation of the operation in an open quantum system, the fidelity of the final state should be generalized as 
\begin{equation}
    f_\mathrm{open}= \langle\mathrm{Tr}\left[\ket{\psi_{\mathrm{o}}}\bra{\psi_{\mathrm{o}}}\mathcal{E}_{T,\delta u_\mathrm{f}}(\ket{\psi_{\mathrm{i}}}\bra{\psi_{\mathrm{i}}})\right]\rangle,
\end{equation}
and this corresponds to the fidelity of the final state in an open quantum system with uncertain Hamiltonians and decoherence. Here, $\mathcal{E}_{T,\delta u_\mathrm{f}}(\cdot)$ describes the completely positive and trace-preserving (CPTP) map between an initial input state $\rho$ and the output state, which is governed by Equ.~\ref{Eq3MaterEq}. As shown in the purple region of Fig.~\ref{fig:Schematic}, the approximate Open-GRAPE algorithm in this work adopts the key ideas of both the Closed-GRAPE and Open-GRAPE algorithms, i.e., it canavoid the calculation of the differentiation and the master equation, while considering the influence of imperfections. We first introduce an approximation that can be applied to $f_{\mathrm{open}}$ to obtain a computationally convenient objective function $J_{\mathrm{open}}$ with $J_\mathrm{open} \approx f_{\mathrm{open}}$.

For an open quantum system, we focus on the demands of optimizing the control pulses for achieving high-fidelity quantum operations under the conditions where noise is weak, i.e., $\kappa_{m}T\ll1$ (thus, $\kappa_{m}\tau\ll1$), and uncertainty is small, i.e., $(\sigma_{f}^{(m)}T)^{2}\ll1$. Consequently, we treat noise as perturbations that can be approximated to the first order (See Supplementary Materials for more details). The final operator corresponding to a single occurrence of the Lindblad jump leads to
\begin{align}
\rho_{L, N}= \tau\sum_{j} U_{j+1 \rightarrow N}\mathscr{L}\left(\rho_{j}\right)U_{j+1 \rightarrow N}^{\dagger},
\end{align}
and we introduce the objective function term due to decoherence
\begin{align}
J_{\mathrm{d}}= \mathrm{Tr}\left(\ket{\psi_{\mathrm{o}}}\bra{\psi_{\mathrm{o}}}\rho_{L, N}\right).
\end{align}
For the parameter uncertainty, the final state corresponding to the  $1$-st order uncertain parameters term after the ensemble average over the random variable $\{\delta u_{\mathrm{f}}^{(m)}\}$ is 
\begin{align}
\overline{\rho_{f, N}}=&-\sum_m(\sigma _{f}^{(m)}\tau) ^2(\sum_{i=2}^{N}{\sum_{j=1}^{i-1}{U_{i+1\rightarrow N}}}H^{(m)}_\mathrm{f}U_{j+1\rightarrow i}H^{(m)}_\mathrm{f}\rho _{j}U_{j+1\rightarrow N}^{\dagger} 
\notag\\
&-U_{j+1\rightarrow N}H^{(m)}_\mathrm{f}\rho _{j}U_{j+1\rightarrow i}^{\dagger}H^{(m)}_\mathrm{f}U_{i+1\rightarrow N}^{\dagger})+h.c.,
\end{align}
where $h.c.$ is the Hermitian conjugate of the former part. The corresponding objective function term induced by this uncertainty is
\begin{align}
    J _{\mathrm{f}}= \mathrm{Tr}(\left|\psi_{\mathrm{o}}\right\rangle \left\langle \psi_{\mathrm{o}}\right|\overline{\rho_{f,N}}).
\end{align}
The detailed deduction and the analytical gradient are shown in Supplementary Materials. Finally, the objective function of the approximate Open-GRAPE algorithm under the combination of these terms is 
\begin{equation}\label{Equ_11:JOpen}
    J_{\mathrm{open}}=J_{\mathrm{close}}+J_{\mathrm{d}} +J_{\mathrm{f}}.
\end{equation}

The goal of the optimization is to maximize this objective function. However, as the fidelity of many current systems approaches 1, it is more intuitive to use the infidelity as $\widetilde{f}_\mathrm{open}=1-\sqrt{f_\mathrm{open}}$ for better evaluation. Therefore, in the subsequent optimization, we equivalently minimize the following objective function
\begin{equation}
    \Phi_{\mathrm{close/open}}=1-\sqrt{J_{\mathrm{close/open}}}~.
\end{equation}
This corresponds to the infidelity commonly adopted by many experimental groups at present.

The approximate Open-GRAPE algorithm in this work inherits the advantage of calculating the gradients using multiple trajectories associated with the extrinsic errors, thus also circumventing the differentiation. Here, we only use fidelity as an example to introduce the algorithm, but we think it is important to extend the algorithm to other figures of merit such as the trace distance~\cite{nielsen2010quantum} and the squared Euclidean distance~\cite{bergholm2019optimal,petruhanov2023grape}.

\begin{figure*}
\begin{centering}
\includegraphics[scale=0.53]{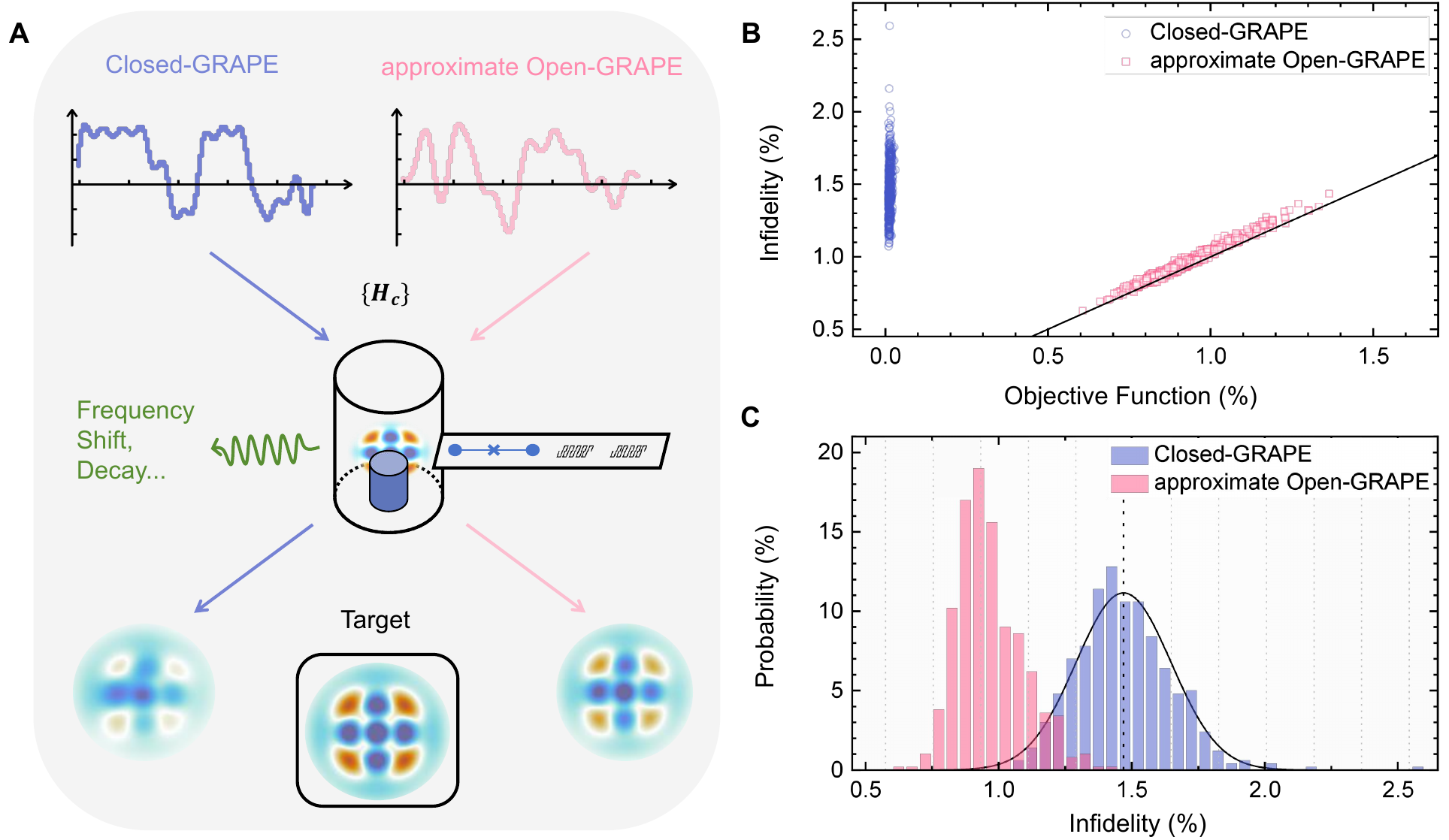}
\par\end{centering}
\caption{\label{fig:Fig2} \textbf{The numerical simulation results.}  (\textbf{A}) The schematic of the numerical simulation. The upper part displays the control pulses obtained by the Closed-GRAPE algorithm (blue) and the approximate Open-GRAPE algorithm (red). The middle part is a superconducting circuit with a three-dimensional cavity and a transmon qubit, where the green line represents the dominant extrinsic noise. The black box at the bottom represents a typical Wigner function of the target quantum state, while the blue and red arrows point to the Wigner functions of the quantum states prepared by the respective pulses from the Closed-GRAPE and approximate Open-GRAPE algorithms. 
(\textbf{B}) The infidelity $\widetilde{f}_{\mathrm{open}}$ of $500$ Closed-GRAPE pulses (blue) and their corresponding approximate 
 Open-GRAPE pulses (red). The objective functions for the Closed-GRAPE and approximate  Open-GRAPE algorithms are $\Phi_\mathrm{close}$ and $\Phi_\mathrm{open}$ (see the main text), respectively. The black line represents the reference line where the objective function in the algorithm is equal to the infidelity considering perturbations. It can be seen that all the red points (from approximate  Open-GRAPE) closely align with the reference line, while the blue points (from Closed-GRAPE) exhibit significant deviations.
(\textbf{C}) The distribution of infidelity corresponding to the data shown in (B). The black line is a Gaussian distribution fitted to the data from the Closed-GRAPE algorithm. The dark gray vertical dashed line indicates the average value, while the distance between the gray dashed lines is the standard deviation $\sigma$ of the Gaussian distribution. In contrast, the data from the approximate  Open-GRAPE shows notably lower average infidelity.}
\end{figure*}

\noindent \textbf{Numerical Simulation}{\large\par}

\noindent In this section, we evaluate the performance of the approximate Open-GRAPE algorithm using a superconducting quantum circuit as an illustrative example, employing comprehensive numerical simulations. As shown in Fig.~\ref{fig:Fig2}A, the physical system consists of a three-dimensional microwave cavity serving as a high-quality storage for quantum states and a transmon serving as the ancillary qubit to provide the necessary non-linearity for the implementation of quantum operations. It is worth noting that such a superconducting system is one of the leading quantum information processing platforms~\cite{cai2021bosonic}, which has shown high operation fidelity for beating the break-even point of quantum error correction~\cite{ni2023beating,sivak2022real}. The robust and optimal control of such a practical open quantum system is urgent for further improving the operation fidelity and realizing the ultimate goal of fault tolerance.

\begin{figure}
\begin{centering}

\includegraphics[scale=0.53]{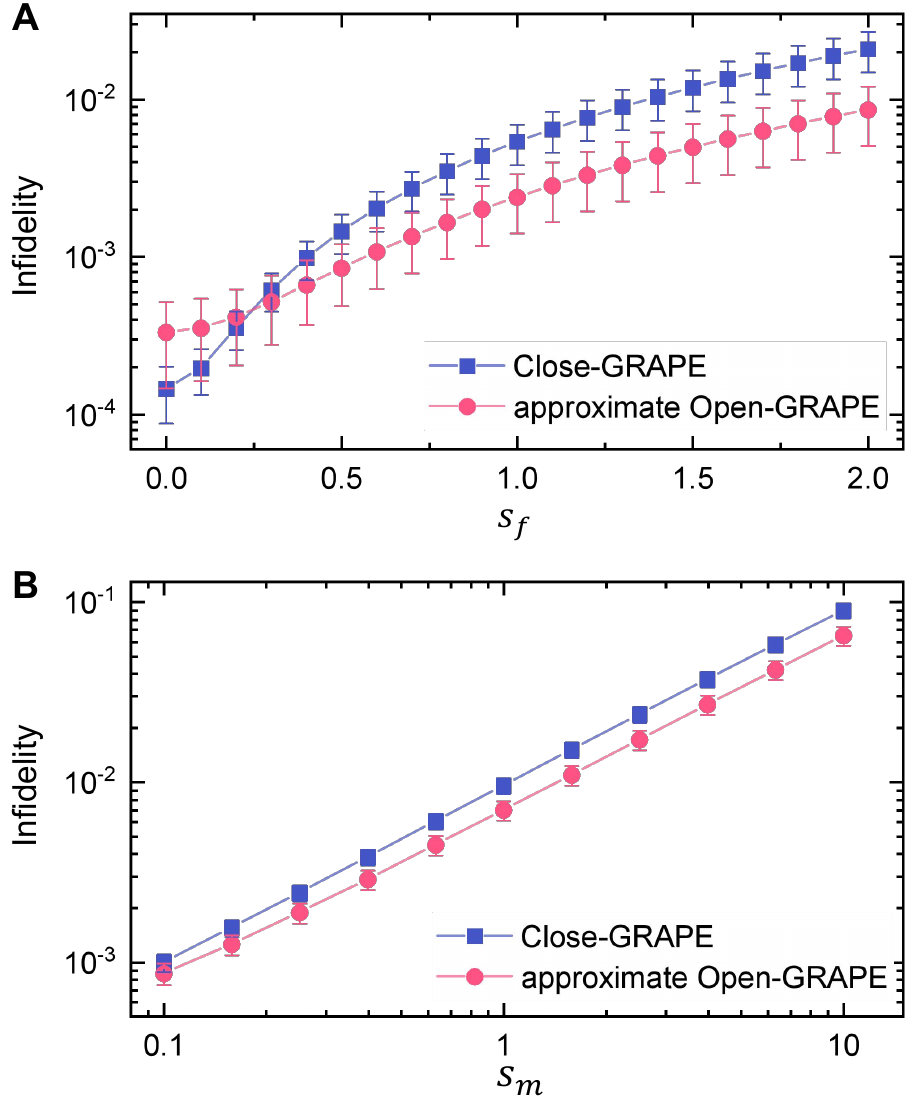}
\par\end{centering}
\caption{\textbf{The performance of the algorithms with varying noise strength.} (\textbf{A}) The average infidelity as a function of the scaling $s_f$ of uncertainty in the Hamiltonian without decoherence noise. (\textbf{B}) The average infidelity as a function of the scaling $s_m$ of decoherence noise strength without Hamiltonian uncertainty. The average infidelities are calculated with $60$ Closed-GRAPE pulses (blue) and their corresponding approximate Open-GRAPE pulses (red). The error bars for the blue points are smaller than the marker size, making them difficult to distinguish visually.}
\label{Fig:RobustNoise}
\end{figure}

The drift Hamiltonian of the bare system in the interaction picture, i.e., a qubit dispersively coupling to a cavity with a cross-Kerr interaction, reads
\begin{align}\label{Equ_13:Num_Hamiltonian}
    H_{0}/\hbar=\frac{\chi}{2}a^{\dagger}a\hat{{\sigma}_z}-\frac{K_2}{2}a^{\dagger}a^{\dagger}aa,
\end{align}
where $a$ and $a^{\dagger}$ are the annihilation and creation operators of the cavity photon, respectively,  $\hat{\sigma}_z$ is the Pauli-Z operator of the transmon qubit,  $\chi$ is the cross-Kerr coupling strength, and $K_2$ is the self-Kerr coefficient of the cavity. Other higher-order interactions can be neglected. Universal quantum operations are realized through coherent microwave drives on both the transmon and the cavity~\cite{heeres2017implementing,hu2019quantum}, where the corresponding control Hamiltonian includes
\begin{align}\label{Equ_14:ControlHamiltonian}
H_{\mathrm{c}}^{(1)}=&  \hat{\sigma}_x, \quad \quad 
H_{\mathrm{c}}^{(2)}=  \hat{\sigma}_y, \notag\\
H_{\mathrm{c}}^{(3)}=&  a+a^{\dagger}, \quad
H_{\mathrm{c}}^{(4)}=  i\left(a-a^{\dagger}\right).
\end{align}
Here $\hat{\sigma}_x$ and $\hat{\sigma}_y$ are the Pauli-X and Pauli-Y operators of the transmon qubit, respectively. 

For a practical experimental system, we consider four types of dominant  perturbations in our numerical simulation, as shown by the green arrow in Fig.~\ref{fig:Fig2}A. For both the cavity and the transmon qubit, the decoherence terms, i.e., $L_{1}=\hat{\sigma}_-$ and $L_2=a$, induced by the decay, and the parameter fluctuations due to frequency shifts, i.e., $H_{f,1}=a^{\dagger}a$ and $H_{f,2}=\hat{\sigma}_z$, are all included. The detailed parameters of the drift and control Hamiltonians as well as the perturbations are shown in the Methods. 

As an important example, we show the result of the encoding process of a binomial code~\cite{michael2016new,hu2019quantum} in the cavity with $|0_B\rangle=(|0\rangle+|4\rangle)/\sqrt{2}$ and $|1_B\rangle=|2\rangle$.  This process transfers the information from the transmon qubit to the cavity encoded with the binomial code, i.e., $\alpha|g\rangle+\beta |e\rangle \rightarrow \alpha|0_B\rangle+\beta |1_B\rangle$. This can be realized with two constraints in the optimization algorithm, and details are shown in the Methods. For the Closed-GRAPE algorithm, different random pulses of the control Hamiltonian are applied as the initial pulses that need to be optimized in the algorithm iteratively until convergence. To better illustrate the differences between the two algorithms and accelerate the numerical simulation, we use the result pulses from the Closed-GRAPE algorithm as the initial inputs for the corresponding approximate Open-GRAPE algorithm optimization. 

The results of $500$ independently optimized pulses are presented in Fig.~\ref{fig:Fig2}B, with the optimized pulses being numerically evaluated via the Lindbald master equation in Equ.~\ref{Eq3MaterEq}. The master equation provides a virtual open quantum system to validate the performance of the target operations with the infidelity $\widetilde{f}_{\mathrm{open}}$. As expected, the pulses provided by Closed-GRAPE exhibit the objective function $\Phi_\mathrm{close}$ that is very close to $0$, while their practical infidelity $\widetilde{f}_{\mathrm{open}}$ is distributed along the $y$-axis with an average value of $1.47\%$ and a standard deviation of $0.18\%$. In contrast, the pulses by approximate Open-GRAPE exhibit objective functions $\Phi_\mathrm{open}$ that deviate from the ideal value of $0$, but they show consistency with the infidelities $\Phi_\mathrm{open} \approx \widetilde{f}_\mathrm{open}$. This agreement indicates the effective evaluation of the gate performance of the open quantum system by our objective function in Equ.~\ref{Equ_11:JOpen}. Benefiting from the good objective function in approximate Open-GRAPE, we find that the average infidelity is improved by $34\%$ to $0.97\%$ and their deviation is reduced to $0.12\%$.

In Fig.~\ref{fig:Fig2}C, the infidelity distribution from the Closed-GRAPE closely resembles a Gaussian distribution. According to this distribution, pulses located beyond three standard deviations from the mean, i.e., pulses with infidelity below $0.93\%$, can be approximately estimated with a probability of $0.135\%$ (0 out of 500 pulses in our simulation). However, the approximate Open-GRAPE yields these pulses with a probability of $46.2\%$ (231 out of 500 pulses) in this simulation, showing an enhancement of more than $340$ times in yield, or more than two orders of magnitude. During the optimization process with multiple parameters, it is easy to converge to local optima. Therefore, in most experiments where necessary prior knowledge is lacking, multiple different random initial pulses are needed for optimization. Then, we can select the one with the best performance. Intuitively, yield represents the probability of generating high-quality pulses. The higher the yield, the fewer trials are required to generate a high-quality pulse on average.

A pronounced asymmetry is shown in the distribution corresponding to the approximate Open-GRAPE, which indicates that the optimized results are close to the lower bound of achievable infidelity in this situation, as opposed to the symmetric Gaussian-like distribution observed in Closed-GRAPE. It is noteworthy that approximate Open-GRAPE achieves a pulse with the best infidelity of $0.63\%$, a value well predicted by our objective function, positioning it around $5\sigma$ in the Gaussian distribution of the Closed-GRAPE, as shown in Fig.~\ref{fig:Fig2}C. Notably, the Open-GRAPE algorithms can also achieve a similar distribution~\cite{schulte2011optimal} but with higher computational complexity due to its precise solution of the system dynamics.

To further demonstrate the robustness of the approximate Open-GRAPE algorithm, we conduct separate optimization for uncertain Hamiltonians and decoherence noise under a particular perturbation strength ($\{\sigma_f\},\{\kappa_m\}$). The robustness is tested by the Master equation with scaled noise strengths $\{s_f\sigma_f\}$ and $\{s_m\kappa_m\}$, where $s_f$ and $s_m$ are scaling factors, and the results are presented in Fig.~\ref{Fig:RobustNoise}. 
As shown in Fig.~\ref{Fig:RobustNoise}A, the approximate Open-GRAPE outperforms the Closed-GRAPE in most parameter regions. Noting that although  the approximate Open-GRAPE is optimized for $s=1$, its performance is better even when the parameter uncertainty tends to vanish. Since the Closed-GRAPE is optimized for $s=0$, its performance is only slightly better than the approximate Open-GRAPE when $s\approx0$. Similarly, the robustness of the approximate Open-GRAPE against decoherence is tested and the result is illustrated in Fig.~\ref{Fig:RobustNoise}B, which shows a similar behavior compared with the Closed-GRAPE. These results indicate the robustness of the approximate Open-GRAPE algorithm to the noise parameters. Even when the parameter uncertainties and the decoherence rates are not precisely calibrated and may shift during experiments, the approximate Open-GRAPE algorithm can consistently provide a reliably improved performance. 

Notably, the numerical pulse sequences generated by the approximate Open-GRAPE algorithm exhibit similar behaviors to those produced by traditional dynamical decoupling techniques~\cite{lidar2014review,barnes2022dynamically,zeng2018general} in resisting the negative effects of uncertain parameters. Although the approximate Open-GRAPE algorithm cannot provide a physical picture from an analytical perspective, it can easily handle a wider variety of noise types and more complicated quantum operations. These details can be found in Supplementary Materials.

\smallskip{}

\noindent \textbf{Experiment}{\large\par}

\begin{figure*}
\begin{centering}
\includegraphics[scale=0.54]{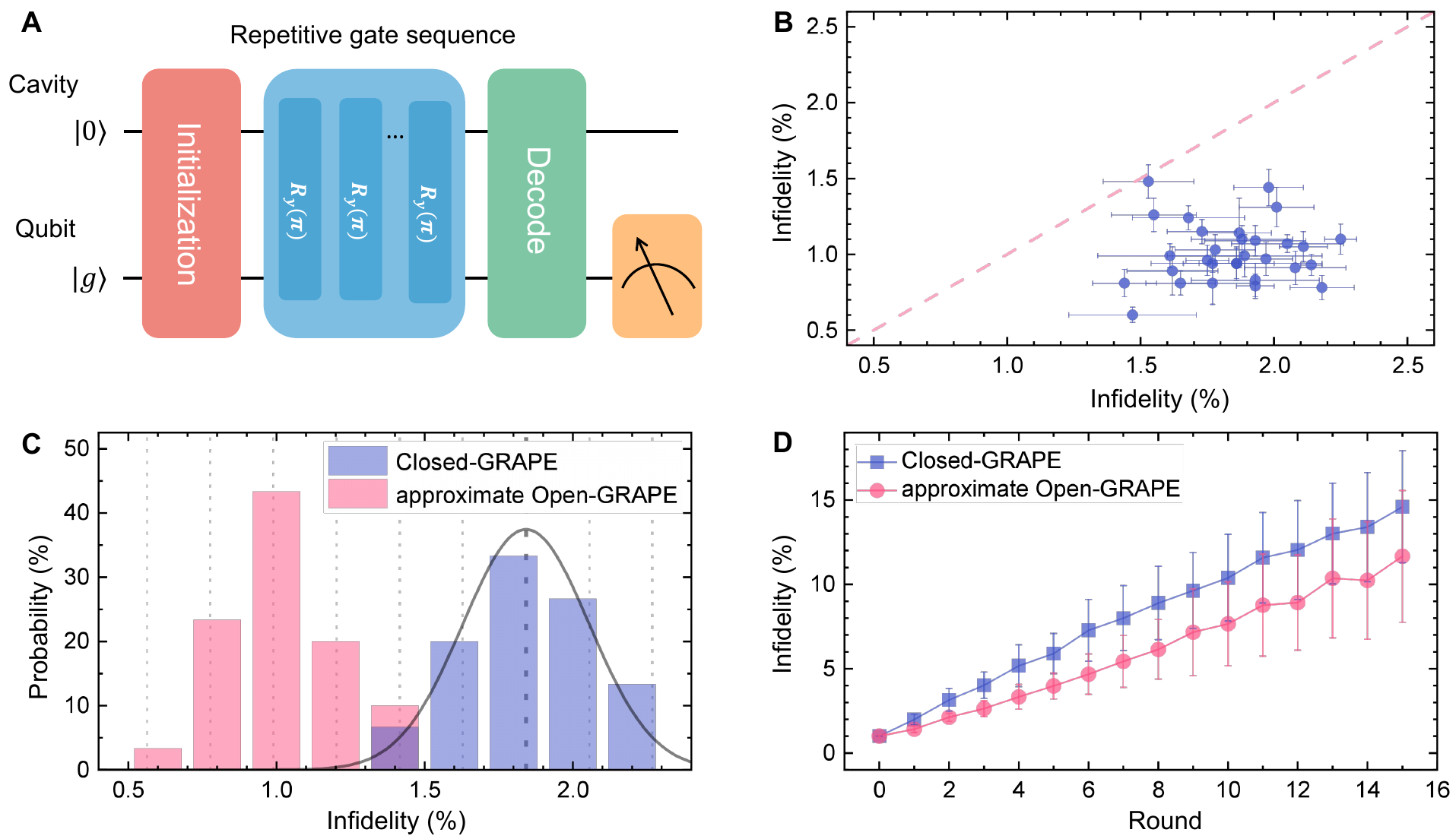}
\par\end{centering}
\caption{\label{fig:Fig4_Experiment}\textbf{The experimental results. }
(\textbf{A}) Quantum circuit for the experiment.
(\textbf{B}) Scatter diagram of infidelities in the initialization experiment with $30$ randomly chosen initial pulses. The horizontal and vertical axes represent the outcomes after optimization with the Closed-GRAPE algorithm and the subsequent refinement through the approximate Open-GRAPE algorithm, respectively. The red dashed line indicates the boundary where the infidelity remains the same with the two algorithms, and the lower half-space is the region where approximate Open-GRAPE is advantageous.
(\textbf{C}) The distribution of infidelity corresponding to the data shown in (B). The black line is a Gaussian distribution fitted to the data from the Closed-GRAPE algorithm. The dark gray vertical dashed line indicates the average value and the distance between the gray dashed lines is the standard deviation $\sigma$ of the Gaussian distribution. 
(\textbf{D}) Infidelity versus the number of repetitive rotation gates. 13 random initial pulses are chosen in this experiment. The red (blue) line is the average infidelity corresponding to the approximate Open-GRAPE (Closed-GRAPE) algorithm. The infidelity deviations for different initial pulses are reflected by the error bars.}
\end{figure*}

\noindent The performance of the approximate Open-GRAPE and Closed-GRAPE algorithms is further verified experimentally with a superconducting quantum circuit, as also studied in the aforementioned numerical simulations. The system consists of a three-dimensional cavity with a high-quality factor of $Q=4.9\times10^{7}$ and an ancillary transmon qubit.  Our practical experimental Hamiltonians are the same as Equs.~\ref{Equ_13:Num_Hamiltonian} and ~\ref{Equ_14:ControlHamiltonian}, with the calibrated $\chi/2\pi=1.00~\mathrm{MHz}$ and $K_2/2\pi=1.415~\mathrm{kHz}$. The dominant errors are the decoherence noise including the decay ($L_1=\hat{\sigma}_{-}$, $T_1=110~\mu\mathrm{s}$) and dephasing ($L_2=\hat{\sigma}_{z}$, $T_2=130~\mathrm{\mu s}$) of the transmon qubit and the relaxation ($L_3=a$, $T_1=1300~\mu \mathrm{s}$) of the storage cavity. The experimental control is implemented with microwave pulses to coherently drive the cavity and the transmon via arbitrary waveform generators (AWGs). The pulse shapes are numerically optimized by the GRAPE algorithms with a step size of $\tau=2~\mathrm{ns}$. The detailed experimental setup, wiring, initialization, pulse generation, and readout are the same as previous experiments~\cite{cai2023protecting}, and are also provided in the Methods. 

With the GRAPE algorithm, the encoding, gates, and decoding of logical qubits based on the binomial codes in the cavity can be realized. All the operations on the logical qubits can be optimized by either the Closed-GRAPE or the approximate Open-GRAPE on-demand. In our experiments, the duration of the initialization and the decoding operations are set to $2~\mathrm{\mu s}$, while the duration of the logical $R_y(\pi)$ gate is $3~\mathrm{\mu s}$. Here, $R_y(\pi)$ denotes a $\pi$ rotation gate around the $y$-axis in the Bloch sphere of the logical qubit. To characterize the performances of the operations based on the optimized controls, we first directly characterize the simplest quantum circuit of a logical qubit consisting of only encoding and decoding, and then the logical $R_y(\pi)$ gate is characterized by repetitively implementing the gate. The control sequence is shown in Fig.~\ref{fig:Fig4_Experiment}A.

For the first experiment, we initialize the logical state to $|+_B \rangle =(|0_B\rangle+|1_B\rangle)/\sqrt{2}$ and then decode the logical state to the transmon for characterization. The constraints in the optimization of the encoding and decoding operations and the detailed parameters are shown in the Methods. The performance of the two operations is characterized by a state tomography of the transmon qubit. From the tomography result, the density matrix of the transmon qubit can be reconstructed numerically and the  infidelity can be calculated. The details of the tomography process can be referred to the Supplementary Materials.
The infidelity of this experiment is shown in Fig.~\ref{fig:Fig4_Experiment}B, with each approximate Open-GRAPE pulse optimized based on the resulting pulse of Closed-GRAPE. It is evident that the approximate Open-GRAPE improves the average infidelity from  $1.84\% \pm 0.21\%$ to $1.01\% \pm 0.20\%$, showing a relative improvement of about $45\%$ in the infidelity. Furthermore, the variance of infidelity along the horizontal axis is generally larger than that along the vertical axis, and this implies that the pulses obtained by the approximate Open-GRAPE algorithm exhibit greater robustness to noise parameters. 

In Fig.~\ref{fig:Fig4_Experiment}C, the infidelities of the Closed-GRAPE algorithm show symmetric Gaussian-like distributions, excellently agreeing with the numerical results in Fig.~\ref{fig:Fig2}C. According to the Gaussian distribution, a probability of $0.135\%$ is predicted to obtain a control pulse realizing an infidelity below $1.20\%$ (three standard deviations below the mean value), while no pulse actually ($0$ out of $30$ pulses in our experiments) achieves this goal. In contrast, a probability of $83.3\%$, i.e., $25$ out of $30$ pulses, is achieved by the approximate Open-GRAPE in our experiments with an ultra-low infidelity of $0.60\%$. These results imply that the main perturbations are accounted for and their impact on quantum control is suppressed. We also note that all parameters in our experiments are not perfectly calibrated and might also fluctuate during experiments, which is notably different from the numerical simulations. Nevertheless, the results demonstrate the robustness of our algorithm. 

To demonstrate the advantage of the approximate Open-GRAPE algorithm in deeper and more complicated circuits, we also design the repetitive logical $R_y(\pi)$ gate sequence as shown in Fig.~\ref{fig:Fig4_Experiment}A. Similarly, the storage cavity is first initialized in the $ |+_B \rangle $ state. Then the logical $R_y(\pi)$ gates are implemented on the state with $M$ repetitions. Eventually, the decoding and tomography process are performed to obtain the infidelity of the final state. It is worth noting that the ideal final state is $|+_B\rangle$ or $|-_B\rangle$ depending on whether $M$ is even or odd, respectively. Figure~\ref{fig:Fig4_Experiment}D shows the average infidelity of $13$ random initial pulses as $M$ increases. From the linear fitting of the infidelity curve of each initial pulse, the infidelity for the logical $R_y(\pi)$ gate can be determined. The average infidelity is $0.89\% \pm 0.21\%$ for the Closed-GRAPE algorithm and $0.72\% \pm 0.28\%$  for the approximate Open-GRAPE algorithm. This shows that the pulses further optimized in the approximate Open-GRAPE algorithm exhibit a relative improvement of about $19\%$ in infidelity. Furthermore, the optimal pulses generated from the approximate Open-GRAPE algorithm demonstrate the lowest infidelity of $0.44\% \pm 0.01\%$. The performance of the $13$ pulses is detailed in the Supplementary Materials. 

\begin{figure}[t]
\begin{centering}
\includegraphics[scale=0.55]{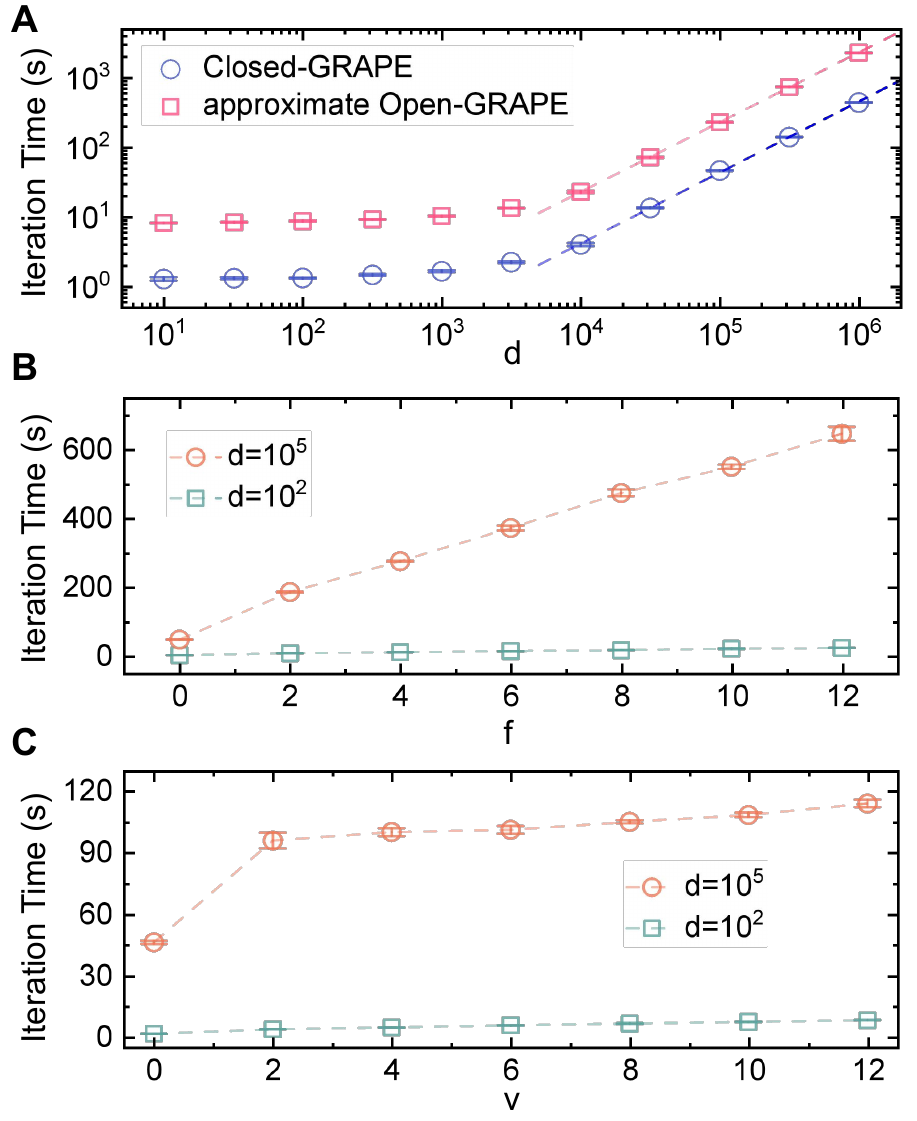}
\par\end{centering}
\caption{\label{fig:CalTime}\textbf{The computational complexity of the algorithms.}  (\textbf{A}) The variation of the iteration time with an increasing state dimension $d$. The blue and red points represent the data of the Closed-GRAPE algorithm and the approximate Open-GRAPE algorithm with $f=2$ and $v=2$. The blue and red dashed lines show the scaling of $3.38\times10^{-4}\times d^{1.02}$ and $2.18\times10^{-3}\times d^{1.00}$, respectively, fitted from the rightmost five blue or red data points. (\textbf{B})  The variation of the iteration time with an increasing number of uncertain Hamiltonians $f$. (\textbf{C}) The variation of the iteration time with an increasing number of decoherence noise sources $v$. The green points and the orange points represent situations with a matrix dimension of $10^2$ and $10^5$, respectively.}
\end{figure}

\smallskip{}
\noindent \textbf{Computational Complexity}{\large\par}
The complexity of the Closed-GRAPE and approximate Open-GRAPE optimization algorithms is important for practical applications. Although the convergence of the algorithms is determined by the specific problem, constraints, and initial guesses, we can quantitatively compare their complexity through the time consumption of calculating the gradient during each iteration in the optimization. 

When the Hilbert space dimension $d$ of the system is large, the most time-consuming process in the GRAPE algorithm is the calculation of the unitary matrix $\{U_j\}$ related to the forward and backward propagations. This is due to the fact that the computational complexity of the matrix exponential of a unitary operator $U=\mathrm{exp}\left(-i H \tau \right)$ from the Hamiltonian $H$ is $\mathcal{O}(d^3)$, while the rest part remains $\mathcal{O}(d^2)$. To avoid this, we apply the matrix-vector exponential approach as outlined in Ref.~\cite{abdelhafez2019gradient} and the similar idea can also be found in the earlier works~\cite{kuprov2007polynomially,kuprov2008polynomially}. The key idea is to calculate the propagation of the state, e.g., $\left|\psi\right\rangle$, using the Taylor expansion of the unitary operator as 
\begin{equation}
U\left|\psi\right\rangle \approx \sum_{m=0}^{n_\mathrm{Taylor}}\left|\phi_m\right\rangle,
\end{equation}
where $\left|\phi_0\right\rangle=\left|\psi\right\rangle$ and $\left|\phi_m\right\rangle=\frac{-i\tau}{m}H\left|\phi_{m-1}\right\rangle$ are terms related to different orders, and $n_\mathrm{Taylor}$ is the truncation order of the Taylor expansion. With this method, the computational complexity of the gradient is demonstrated in the Supplementary Materials and the results show that the complexity is $\mathcal{O}[(2n_\mathrm{Taylor}+n_\mathrm{control})Nd^2]$ for the Closed-GRAPE algorithm and $\mathcal{O}\{[(2v+2f+13)n_{\mathrm{control}}+2(f+3)n_\mathrm{Taylor}+2(f+2v+1)] N d^2\}$ for approximate Open-GRAPE algorithm. Here, $n_\mathrm{control}$ is the number of control Hamiltonians to be optimized, $v$ is the number of decoherence noise sources, and $f$ is the number of uncertain Hamiltonians. Comparing the two approaches, our approximate Open-GRAPE algorithm consumes more time but only by a constant factor. For example, when $v=2$, $f=2$, $n_\mathrm{control}=4$, and $n_\mathrm{Taylor}=20$, the complexity of the approximate Open-GRAPE algorithm is only $6.77$ times that of the Closed-GRAPE. This scaling is acceptable for practical experimental systems in which the main uncertain parameters and decoherence channels are limited to a small number. In this perspective, our approximate Open-GRAPE algorithm outperforms previous attempts to optimal control of open quantum systems based on the GRAPE algorithm~\cite{boutin2017resonator,goerz2015optimizing}, whose computational complexity is at least $\mathcal{O}(d^3)$ when performing the optimization with the $d \times d$-dimensional density matrix representation. 

The computational complexity is also numerically tested by recording the duration for each iteration of the GRAPE algorithms. Figure~\ref{fig:CalTime}A directly compares the time consumption of the Closed-GRAPE and approximate Open-GRAPE algorithms as the system dimensions increase. The data illustrate a trend that is lower than quadratic scaling and approaches linearity when $d$ is large, which may be attributed to the sparsity of the Hamiltonian matrices. However, due to the foundational overhead of computer operations such as array initialization, the iteration time becomes relatively higher than expected when $d$ is small, resulting in an inflection point around $d=10^4$ in the figure.
The increasing time consumption with the considered uncertain parameters and noise terms are also numerically investigated, as shown in Figs.~\ref{fig:CalTime}B and \ref{fig:CalTime}C. As expected, the calculation time shows a nearly linear increase with both $f$ and $v$ for $d=10^2$ below the inflection point and $d=10^5$ above the point. These numerical results validate the computational efficiency of our algorithm for practical applications. For example, in the qubit-cavity model investigated earlier with $f=2$ and $v=2$ perturbations, our results imply that the typical iteration duration for the $2\times5000$-dimension problem is about $22$ seconds. Furthermore, the results also show that optimizing open quantum system controls with dimensions $d\approx 10^6$ becomes achievable on personal computers, equivalent to handling roughly $20$ qubits.

\smallskip{}

\noindent \textbf{DISCUSSION}{\large\par}
In summary, a numerically efficient GRAPE algorithm for optimal and robust control of open quantum systems is proposed and experimentally verified. This approximate Open-GRAPE algorithm, instead of offering entirely new control solutions that the Closed-GRAPE could not uncover, shows a higher probability of finding the potential control pulses that suppress the impact of perturbations and are robust against parameter uncertainties. Both simulation and experimental results affirm the computational efficacy of our algorithm, showcasing a 340-fold increase in the probability of generating high-performance pulses, while maintaining a modest linear increase in complexity for calculating the gradient in each iteration.

This new algorithm allows us to approach the practical lower bound of infidelities when controlling open quantum systems in practice, promising much better performance based on current system parameters. For example, while the average infidelity using the Closed-GRAPE algorithm can be as large as $1.84\%$, the best optimization results achieved with the approximate Open-GRAPE can reach $0.60\%$ as demonstrated in the experiment. Furthermore, our algorithm provides an effective means to estimate the minimal physical resource requirement to achieve a target operation precision, which is crucial for evaluating the hardware needs to realize fault tolerance thresholds and quantum supremacy. Our results imply that the stringent demands on the parameters of the system, such as $T_1$ and $T_2$ of the qubits, can be relaxed by approaching the lower bound of infidelities with the approximate Open-GRAPE algorithm.

High gate fidelity is immediately important for applications in the noisy intermediate-scale quantum era, as well as for the exploration of quantum error correction and fault-tolerant techniques. Even a marginal improvement in fidelity can  substantially increase achievable quantum circuit depth or enable quantum error correction codes to surpass the break-even point. Hence, it is anticipated that the approximate Open-GRAPE algorithm can contribute substantially to these applications. Additionally, our algorithm is generally applicable to open quantum systems and thus can be easily extendable to other quantum platforms, including Rydberg atoms~\cite{omran2019generation} and trapped ions~\cite{figgatt2019parallel,yang2022fast}. Furthermore, our treatment of perturbations in the GRAPE algorithm can be further generalized to applications beyond gate operations~\cite{harrington2022engineered}, such as quantum metrology~\cite{wang2021quantum}, quantum state resetting~\cite{magnard2018fast}, and quantum simulation~\cite{daley2022practical}. Additionally, to enhance the practical applicability of the algorithm, we need to further improve its yield by addressing challenges such as avoiding local optima. Finally, to further enhance the performance of quantum systems, we need to improve the algorithm to effectively handle low-frequency and non-Markovian noise~\cite{koch2022quantum}.

\clearpage

\noindent \textbf{\large{}METHODS}{\large\par}

\noindent \textbf{Constraints of the Algorithm}

In the numerical simulation, the encoding process of the binomial code can be realized through the following constraints:
\begin{align}
\left| \psi _{\mathrm{i}}^{1} \right. \rangle &=\left| g \right. \rangle \otimes \left| 0 \right. \rangle,
&
\left| \psi_{\mathrm{o}}^{1}\right\rangle  &=\left| g \right. \rangle \otimes |1_B\rangle;
\notag\\
\left| \psi _{\mathrm{i}}^{2} \right. \rangle &=\frac{\left| g \right. \rangle +\left| e \right. \rangle}{\sqrt{2}}\otimes \left| 0 \right. \rangle,
&
\left| \psi _{\mathrm{o}}^{2} \right. \rangle &=\left| g \right. \rangle \otimes |+_B\rangle;
\end{align}
where $|\pm_B\rangle=(|0_B\rangle \pm |1_B\rangle)/\sqrt{2}$, and $|0_B\rangle=(|0\rangle+|4\rangle)/\sqrt{2}$ and $|1_B\rangle=|2\rangle$ are the two code basis states. The unitary evolution satisfying these constraints can transfer the state from $(\alpha|g\rangle+\beta |e\rangle) \otimes |0\rangle$ to $ |g\rangle \otimes (\alpha|0_B\rangle+\beta |1_B\rangle)$. Here, no more constraints are set to the evolution in other subspaces and the global phase is also neglected. 

In the experiment, the initialization process is realized with only a single constraint:
\begin{align}
    \left| \psi _{\mathrm{i}}^{1} \right. \rangle&=\left| g \right. \rangle \otimes \left| 0 \right. \rangle
,&
\left| \psi _{\mathrm{o}}^{1} \right. \rangle &=\left| e \right. \rangle \otimes |+_B\rangle.
\end{align}
The constraints on the decoding process are more stringent with:
\begin{align}
\left| \psi _{\mathrm{i}}^{1} \right. \rangle&=\left| e \right. \rangle \otimes |+_B \rangle, 
&
\left| \psi _{\mathrm{o}}^{1} \right. \rangle&=(|g\rangle+|e\rangle)/\sqrt{2}\otimes |0\rangle;
\notag \\
\left| \psi _{\mathrm{i}}^{2} \right. \rangle&=\left| e \right. \rangle \otimes |-_B \rangle, 
&
\left| \psi _{\mathrm{o}}^{2} \right. \rangle&= (|g\rangle-|e\rangle)/\sqrt{2}\otimes |0\rangle ;
\notag \\
\left| \psi _{\mathrm{i}}^{3} \right. \rangle&=\left| e \right. \rangle \otimes |0_B \rangle, 
&
\left| \psi _{\mathrm{o}}^{3} \right. \rangle&= |g\rangle\otimes |0\rangle; 
\notag\\
\left| \psi _{\mathrm{i}}^{4} \right. \rangle&=\left| e \right. \rangle \otimes |1_B \rangle, 
&
\left| \psi _{\mathrm{o}}^{4} \right. \rangle&= |e\rangle\otimes |0\rangle. 
\end{align}
This decoding process can realize arbitrary state transfers from the storage cavity to the ancillary qubit, i.e., from $ |e\rangle \otimes (\alpha|0_B\rangle+\beta |1_B\rangle)$ to $(\alpha|g\rangle+\beta |e\rangle) \otimes |0\rangle$.

The constraints on the logical $R_y(\pi)$ gate neglecting the global phase are : 
\begin{align}
\left| \psi _{\mathrm{i}}^{1} \right. \rangle&= \left| e \right. \rangle \otimes  |+_B \rangle,& 
\left| \psi _{\mathrm{o}}^{1} \right. \rangle &= \left| e \right. \rangle \otimes |-_B \rangle;
\notag\\
\left| \psi _{\mathrm{i}}^{2} \right. \rangle&=\left| e \right. \rangle \otimes |0_B\rangle,& 
\left| \psi _{\mathrm{o}}^{2} \right. \rangle &= \left| e \right. \rangle \otimes |1_B \rangle.
\end{align}

\noindent \textbf{Simulation Details}

In the Numerical Simulation section of the main text, the GRAPE algorithm is computed using the following Hamiltonian in the interaction picture, with the higher-order terms being neglected: 
\begin{align}
    H/\hbar=-\chi a^{\dagger}a\hat{\sigma}_+\hat{\sigma}_--\frac{K_2}{2}a^{\dagger}a^{\dagger}aa.
\end{align}
To effectively demonstrate the optimization efficacy of the approximate Open-GRAPE algorithm, the parameters used in this section are different from those in the real experiment.
For the four control Hamiltonians shown in Equ.\ref{Equ_14:ControlHamiltonian} of the main text, the maximum allowable amplitude during the optimization in the GRAPE algorithm is $50~\mathrm{MHz}$. The cross-Kerr coupling strength is $\chi/2\pi=1.9~\mathrm{MHz}$ and the self-Kerr coefficient of the storage cavity is $K_2/2\pi=8.46~\mathrm{kHz}$.
 
In Figs.~\ref{fig:Fig2}B and \ref{fig:Fig2}C of the main text, the dimensions of the cavity and the transmon qubit are chosen to be $30$ and $2$, respectively. The overall duration time $T=600~\mathrm{ns}$ is equally split into $N=600$ steps. The frequency shift fluctuations are intentionally increased beyond those observed in the real system and are chosen to be $\sigma^{(1)}_\mathrm{f}=\sigma^{(2)}_\mathrm{f}=0.1~\mathrm{MHz}$ for both the storage cavity and the transmon qubit. The corresponding  Hamiltonians are $H^{(1)}_\mathrm{f}=a^\dagger a$ and $H^{(2)}_\mathrm{f}=\hat{\sigma}_+\hat{\sigma}_-$, respectively. Only the relaxation noise of the cavity and the transmon qubit are considered here, i.e., $L_1=a$ and $L_2=\hat{\sigma}_{-}$. Their strengths are $\kappa_1=10~\mathrm{kHz}$ and $\kappa_2=50~\mathrm{kHz}$, respectively.

Calculations of the average infidelity in the Numerical Simulation section are obtained by computing the weighted average infidelity between the target states specified in the constraints and the corresponding noisy final states after evolution. To simulate the influence of the fluctuating Hamiltonian, the uncertain parameters are assumed to follow a binomial distribution. Consequently, the infidelity is averaged over the final states resulting from the evolution with two different Hamiltonians, $H\left(t\right)=H_{0}+\sum_{k}u_{c}^{(k)}\left(t\right)H_{c}^{(k)}+\sum_{m} \sigma_{\mathrm{f}}^{(m)}H^{(m)}_\mathrm{f}$ and  $H\left(t\right)=H_{0}+\sum_{k}u_{c}^{(k)}\left(t\right)H_{c}^{(k)}-\sum_{m} \sigma_\mathrm{f}^{(m)}H^{(m)}_\mathrm{f}$.

In the calculation shown in Fig.~\ref{fig:CalTime} of the main text, the overall duration time $T=100~\mathrm{ns}$ is equally split into $N=100$ steps. The calculation is implemented with an Intel(R) Core(TM) i7-8700 CPU. In Fig.~\ref{fig:CalTime}A, the total dimension $d$ is changed by varying the cavity dimension while keeping the transmon qubit dimension fixed at $2$. Only a single constraint is considered with random initial and target states in the analysis. 

\noindent \textbf{Experimental Setup}

\noindent The experiment is conducted on a superconducting circuit including a three-dimensional cavity with a high-quality factor of $Q=4.9\times10^{7}$ and an ancillary transmon qubit.
The parameters of the Hamiltonians are well calibrated, where the cross-Kerr coupling strength is $\chi/2\pi=1.00~\mathrm{MHz}$ and the first-order self-Kerr coefficient is $K_2/2\pi=1.415~\mathrm{kHz}$. Both the initialization and decoding processes have a duration of $2~\mathrm{\mu s}$, while the logical rotation gate $R_y(\pi)$ has a duration of $3~\mathrm{\mu s}$. The time interval for each step in the GRAPE algorithm is $\tau=~2\mathrm{ns}$, resulting in $N=1000$ and $1500$ steps for the initialization and $R_y(\pi)$ gate, respectively. In the experiment, the pulses are generated from an arbitrary waveform generator (AWG) with a minimum time resolution of $0.4~\mathrm{ns}$.

In the experiment, the dominant error source is the decoherence noise, including the decay ($L_1=\hat{\sigma}_{-}$, $T_1=110~\mathrm{\mu s}$) and dephasing ($L_2=\hat{\sigma}_{z}$, $T_2=130~\mathrm{\mu s}$) of the transmon qubit, and the relaxation ($L_3=a$, $T_1=1300~\mathrm{\mu s}$) of the storage cavity. These extrinsic errors are considered in the optimization process of the approximate Open-GRAPE algorithm. Moreover, to prevent pulse distortion from the AWG, several penalty terms are included in both the Closed-GRAPE and approximate Open-GRAPE algorithms in the experiment~\cite{ofek2016extending}. More details can be referred to the Supplementary Materials. Similar to the approach in the Numerical Simulation section, the pulses optimized using the Closed-GRAPE algorithm are utilized as the initial pulses in the approximate Open-GRAPE algorithm for better comparison.

\vbox{}

\vbox{}

\noindent \textbf{SUPPLEMENTARY MATERIALS} The online version contains Supplementary Materials.






\smallskip{}
\bibliographystyle{Zou}
\bibliography{mycite}


\smallskip{}
\smallskip{}
\noindent \textbf{Acknowledgment:}
We are grateful to Weizhou Cai for helpful discussions. 
\textbf{Funding:}
This work was funded by the National Key R\&D Program (Grant No. 2021YFA1402004 and 2017YFA0304303), the National Natural Science Foundation of China (Grants No. 11925404, 92165209, 92365301,12061131011, 92265210),  Innovation Program for Quantum Science and Technology (Grant No.~2021ZD0300203 and 2021ZD0301203). This work was also supported by the Fundamental Research Funds for the Central Universities and USTC Research Funds of the Double First-Class Initiative.  This work was partially carried out at the USTC Center for Micro and Nanoscale Research and Fabrication.
The numerical calculations in this paper have been done on the supercomputing system in the Supercomputing Center of University of Science and Technology of China.
\textbf{Author contributions:}
Z.-J.C. and C.-L.Z. conceived the idea. Z.-J.C. developed the methods.  H.H. and L.D.S. performed the experiment and analyzed the data with the assistance of J.Z., Z.H., Y.X., and W.W. under the supervision of L.S. Z.-J.C., H.H., L.D.S., C.-L.Z., L.S., and X.-B.Z. wrote the manuscript, with feedback from all other authors. C.-L.Z., L.S., and X.-B.Z. supervised the project.
\textbf{Competing interests:}
The authors declare no competing interests.
\textbf{Data and materials availability:}
All data needed to evaluate the conclusions in the paper are present in the paper or the Supplementary Materials.

\smallskip{}

\clearpage{}

\clearpage{}
\onecolumngrid

\clearpage{}
\onecolumngrid
\renewcommand{\thefigure}{S\arabic{figure}}
\setcounter{figure}{0} 
\renewcommand{\thepage}{S\arabic{page}}
\setcounter{page}{1} 
\renewcommand{\theequation}{S.\arabic{equation}}
\setcounter{equation}{0} 
\setcounter{section}{0}

\clearpage{}

\begin{center}
  \Large\bfseries\makebox[0pt]{Supplementary Information for ``Robust and optimal control of open quantum systems''}\\[1em]
\end{center}

\maketitle

\begin{center}
Zi-Jie Chen$^{1\ast}$, \and
Hongwei Huang$^{2\ast}$, \and
Lida Sun$^{2\ast}$, \and
Qing-Xuan Jie$^{1,3}$, \and
Jie Zhou$^{2}$, \and
Ziyue Hua$^{2}$, \and
Yifang Xu$^{2}$,\\
Weiting Wang$^{2}$, \and
Guang-Can Guo$^{1,3,4}$, \and
Chang-Ling Zou$^{1,3,4\dagger}$, \and
Luyan Sun$^{2,4\dagger}$, \and
Xu-Bo Zou$^{1,3,4\dagger}$\\
\small$^{1}$CAS Key Laboratory of Quantum Information, University of Science and Technology of China, \\
\small Hefei, Anhui 230026, P. R. China.\\
\small$^{2}$Center for Quantum Information, Institute for Interdisciplinary Information Sciences, \\
\small Tsinghua University, Beijing 100084, China.\\
\small$^{3}$CAS Center For Excellence in Quantum Information and Quantum Physics, \\
\small University of Science and Technology of China, Hefei, Anhui 230026, China.\\
\small$^{4}$ Hefei National Laboratory, Hefei 230088, China.\\
\small$^\dagger$Corresponding author. Email:  clzou321@ustc.edu.cn (C.-L. Z.);\\ 
\small luyansun@tsinghua.edu.cn (L. S.); xbz@ustc.edu.cn (X.-B. Z.).\\
\small$^\ast$These authors contributed equally to this work.
\end{center}

\onecolumngrid

\tableofcontents
\clearpage

\section{Algorithm Description}

\subsection{Objective Function}
In this section, we present a detailed derivation of the objective function $J_\mathrm{open}$ in the approximate Open-GRAPE algorithm. This function serves as an approximation of the average fidelity in open quantum systems with uncertain parameters and decoherence noise. The key idea in our analysis is to expand the final state after the noisy evolution perturbatively. The zeroth-order term in this expansion is the noiseless term, which is the same as the objective function in the Closed-GRAPE algorithm. With this term, the first-order term considering uncertain Hamiltonian and decoherence noise can be expressed analytically.

As described in the main text, the objective function is approximate to the weighted average fidelity of the final states  $\{\rho^{j}\left(T\right)\}$ with respect to the target states $\{\left|\psi_{\mathrm{o}}^{j}\right\rangle\}$. Here, $\rho^{j}\left(T\right)=\mathcal{E}_{T,\delta u_\mathrm{f}}(\left|\psi_{\mathrm{i}}^{j}\right\rangle \left\langle \psi_{\mathrm{i}}^{j} \right|)$ is the final state corresponding to the pure initial states $\left|\psi_{\mathrm{i}}^{j}\right\rangle$ after the evolution in an open quantum system. For simplicity, our derivation focuses on a single constraint $\{ \left|\psi_{\mathrm{i}}\right\rangle \mapsto \left|\psi_{\mathrm{o}}\right\rangle \}$, where  $\left|\psi_{\mathrm{i}}\right\rangle$ is the initial state, $\left|\psi_{\mathrm{o}}\right\rangle$ is the target state, and $\rho(T)$ is the corresponding noisy final state. The extension from single to multiple states is relatively straightforward, and we will provide the conclusions directly for the multiple-state case at the end. In this situation with a single constraint, the fidelity of the final state is 
\begin{align}
f_\mathrm{open}=\langle\mathrm{Tr}\left[\ket{\psi_{\mathrm{o}}}\bra{\psi_{\mathrm{o}}} \rho\left(T\right)\right]\rangle,
\end{align} 
where $\langle \cdot \rangle$ denotes the average over the distribution of the random parameters $\{\delta u^{(m)}_\mathrm{f}\}$.
The evolution of the system can be described by the Lindblad master equation~\cite{jacobs2014quantum}:
\begin{align}
    \frac{d}{dt}\rho\left( t\right)=-i \left[ \widetilde{H}\left(t\right),\rho \left(t\right) \right]+ \mathscr{L} \left( \rho \left(t\right) \right).
\end{align}
Here, $\rho\left(t\right)$ is the density matrix of the system, 
\begin{align}\label{Lindblad equation}
\mathscr{L}\left(\rho\right)=\sum_{m}\kappa_{m}\left\{ L_{m}\rho L_{m}^{\dagger}-\frac{1}{2}L_{m}^{\dagger}L_{m}\rho-\frac{1}{2}\rho L_{m}^{\dagger}L_{m}\right\}
\end{align}
is the Lindblad superoperator,  $L_{m}$ is the decoherence noise, and $\kappa_{m}$ is the corresponding strength. $\widetilde{H}\left(t\right)$ is the Hamiltonian that can be written as
\begin{equation}
    \widetilde{H}\left(t\right)=H\left(t\right)+\sum_{m}\delta u_{\mathrm{f}}^{(m)}H_{\mathrm{f}}^{(m)},
\end{equation}
where $H\left(t\right)=H_{0}+\sum_{k}u_{c}^{(k)}\left(t\right)H_{c}^{(k)}$ is the Hamiltonian without parameter uncertainty, $H_{0}$ is the drift Hamiltonian,  $H_{c}^{(k)}$ is the $k$-th control Hamiltonian, and  $u_{c}^{(k)}\left(t\right)$ is the corresponding time-varying amplitude parameter to be optimized. Here, $\{H_{\mathrm{f}}^{(m)}\}$ are the fluctuation terms with uncertain amplitudes $\delta u_\mathrm{f}^{(m)}$ satisfying $\left\langle \delta u_{\mathrm{f}}^{(m)}\right\rangle =0$ and $\left\langle (\delta u_{\mathrm{f}}^{(m)})^{2}\right\rangle =(\sigma_{\mathrm{f}}^{(m)})^{2}$.

In this analysis, the decoherence  noise is assumed to be weak, i.e., $\kappa_{m}T\ll1$ (thus, $\kappa_{m}\tau\ll1$).
Under perturbation expansion, the state becomes $\rho(t)=\rho_{0}(t)+\rho_{1}(t)+\rho_{2}(t)+\cdots$, where the zeroth-order term $\rho_{0}(t)$ corresponding to the evolution with uncertainty parameters but without decoherence is
\[
\frac{d}{dt}\rho_{0}\left(t\right)=-i\left[\widetilde{H}\left(t\right),\rho_{0}\left(t\right)\right],
\]and higher-order terms correspond to the evolution as 
\begin{align}\label{PerturbationExpansion}
\frac{d}{dt}\rho_{n}\left(t\right)=-i\left[\widetilde{H}\left(t\right),\rho_{n}\left(t\right)\right]+\mathscr{L}\left(\rho_{n-1}\left(t\right)\right),
\end{align}
for $n\geq1$.

\subsubsection{Parameter uncertainty}
In this section, we first derive the approximation term corresponding to the parameter uncertainty with $(\sigma^{(m)}_{\mathrm{f}})^{2}T^{2}\ll1$. A closed-form expression for this term without decoherence noise can be obtained through the expansion as 
\begin{align}\label{Dyson}
\rho _{0}\left( T \right) 
&=\rho _{0}\left( 0 \right) -i\int_0^T{\left[\widetilde{H}\left( \tau \right) ,\rho _{0}\left( \tau \right) \right] d\tau} 
\notag\\
&=\rho _{0}\left( 0 \right) -i\int_0^T{d\tau \left[ \widetilde{H}\left( \tau \right) ,\rho _{0}\left( 0 \right) \right]}+\left( -i \right) ^2\int_0^T{d\tau}\int_0^{\tau}{d\tau _1\left[ \widetilde{H}\left( \tau \right) ,\left[ \widetilde{H}\left( \tau _1 \right) ,\rho _{0}\left(0 \right) \right] \right]}+\cdots.
\end{align}
Here, we consider the piece-wise control in which the control amplitude  $u_{c}^{(k)}\left(t\right)=u^{(k)}_{c,j}$  remains constant for $t\in\left(\left(j-1\right)\tau,j\tau\right]$. The unitary operator with uncertain parameters in the $j$-th step is  
\begin{align}\label{propagator}
\widetilde{U}_j&=\mathrm{exp}\left[ -i\tau \left(H_0+\sum_k{u^{(k)}_{c,j}}H^{(k)}_c+\sum_m{\delta}u^{(m)}_\mathrm{f} H^{(m)}_\mathrm{f} \right) \right] \notag\\
&\approx \left( I-i\tau \sum_m{\delta}u^{(m)}_\mathrm{f} H^{(m)}_\mathrm{f}\right) U_j,
\end{align}
where $U_j=\mathrm{exp}\left[ -i\tau \left( H_0+\sum_k{u^{(k)}_{c,j}}H^{(k)}_c \right) \right]$ is the unitary operator without uncertain parameters. Therefore, the state in Equ.~\ref{Dyson} after the $j$-th step is 
\begin{align}
    \rho _{0,j}
    =&\rho_{0} \left( j\tau \right) \\ \notag
    =&\widetilde{U}_{1 \rightarrow j} \rho _0\left( 0 \right) \widetilde{U}_{1 \rightarrow j}^{\dagger}.
\end{align}
Here, we denote the unitary operator from  $j$-th to  $(j+i)$-th step as  $\widetilde{U}_{j \rightarrow j+i}=\widetilde{U}_{j+i}\cdots\widetilde{U}_{j+1}\widetilde{U}_{j}$  ($\widetilde{U}_0=I$ and $\widetilde{U}_{i\rightarrow j}=I$ for $i>j$) for simplicity.

To obtain the fidelity of the final state, it is necessary to  calculate the average fidelity over the uncertain parameters under perturbation, i.e., 
\begin{align}\label{inf_fluc}
&\langle \mathrm{Tr}\left( \left| \psi _{\mathrm{o}} \right. \rangle \left. \langle \psi _{\mathrm{o}} \right|\rho_0 (T) \right) \rangle 
\notag \\
=&\left< \left. \langle \psi _{\mathrm{o}} \right|\widetilde{U}_{1\rightarrow N}\rho_0 (0)\widetilde{U}_{1\rightarrow N}^{\dagger}\left| \psi _{\mathrm{o}} \right. \rangle \right>  
\notag \\
\approx& J_\mathrm{close} +J_\mathrm{f}.
\end{align}
Here,
\begin{align}\label{inf_0th}
    J_\mathrm{close}=\left. \langle \psi _{\mathrm{o}} \right|U_{1\rightarrow N}\rho _0U_{1\rightarrow N}^{\dagger}\left| \psi _{\mathrm{o}} \right. \rangle
\end{align}
is the objective function in the Closed-GRAPE algorithm, and it is equal to the fidelity of the final state with respect to the target state in the closed system.  The influence of uncertain Hamiltonians is included in the following term:
\begin{align}\label{inf_fluc_1}
J_\mathrm{f}
=&- \sum_{m}(\tau\sigma^{(m)}_{\mathrm{f}})^{2}\sum_{j=2}^{N}{\sum_{i=1}^{j-1}{(\left. \langle \psi _{\mathrm{o}} \right|U_{j+1\rightarrow N}H^{(m)}_\mathrm{f} U_{i+1\rightarrow j}H^{(m)}_\mathrm{f} U_{1\rightarrow i}\rho _0 U_{1\rightarrow N}^{\dagger}\left| \psi _{\mathrm{o}} \right. \rangle}}
\notag\\
& -\left. \langle \psi _{\mathrm{o}} \right|U_{i+1\rightarrow N}H^{(m)}_\mathrm{f}U_{1\rightarrow i}\rho _0U_{1\rightarrow j}^{\dagger}H^{(m)}_\mathrm{f}U_{j\rightarrow N}^{\dagger}\left| \psi _{\mathrm{o}} \right. \rangle +c.c.),
\end{align}
where $c.c.$ is the complex conjugate of the former part inside the parentheses.
Here, the derivation of Equ.~\ref{inf_fluc} utilizes expansion of the operator in Equ. \ref{propagator} with respect to $\delta u^{(m)}_\mathrm{f}$. The average value of the first-order term of the uncertain parameter is 0 since $\left\langle \delta u^{(m)}_{\mathrm{f}}\right\rangle =0$. The second term $J_\mathrm{f}$  is the fidelity loss term induced by the uncertain parameters in the approximate Open-GRAPE algorithm, corresponding to the variance of the uncertain parameters $\left\langle (\delta u^{(m)}_{\mathrm{f}})^2\right\rangle =(\sigma^{(m)}_\mathrm{f})^2$. Other higher-order terms related to uncertain parameters are neglected under the small uncertainty approximation $(\sigma^{(m)}_{\mathrm{f}})^{2}T^{2}\ll1$.  To be reminded, the term $J_\mathrm{close}$ is also the objective function in the Closed-GRAPE algorithm in Ref.~\cite{khaneja2005optimal}.

Equation~\ref{inf_fluc_1} can be further expressed in the following form with higher-order uncertain parameter terms neglected. This form can provide a more intuitive representation of the subsequent algorithmic calculations:
\begin{align}\label{inf_fluc_2}
J_\mathrm{f}\approx & \sum_{m}\left[ \left. \langle \psi _{\mathrm{o}} \right|\left( I-i\tau \sigma^{(m)}_\mathrm{f} H^{(m)}_\mathrm{f} \right) U_{N}\cdots\left( I-i\tau \sigma^{(m)}_\mathrm{f} H^{(m)}_\mathrm{f} \right) U_2\left( I-i\tau \sigma^{(m)}_\mathrm{f} H^{(m)}_\mathrm{f} \right) U_1\left| \psi_{\mathrm{i}} \right. \rangle \times c.c. \right] 
\notag\\
&-\sum_{m}( \langle \psi _\mathrm{o} |U_{1\rightarrow N}\left| \psi_{\mathrm{i}}\right. \rangle \times c.c.)
\notag\\
& +\sum_{m}\left( i\tau \sigma^{(m)} _\mathrm{f}\sum_{i=1}^{N}{\left. \langle \psi _\mathrm{o} \right|U_{i+1\rightarrow N}H^{(m)}_\mathrm{f}U_{1\rightarrow i}\left| \psi _{\mathrm{i}} \right. \rangle \left. \langle \psi _{\mathrm{i}} \right|U_{1\rightarrow N}^{\dagger}\left| \psi _\mathrm{o} \right. \rangle}+c.c. \right) .
\end{align}
This formulation can be understood as substituting $\delta u^{(m)}_{\mathrm{f}}$ with $\sigma^{(m)}_{\mathrm{f}}$ in Equ.~\ref{inf_fluc} and then subtracting the resulting first-order and zeroth-order terms.
Finally, the objective function of the multiple-state-transfer constraints $\{ \left|\psi_{\mathrm{i}}^{j}\right\rangle \mapsto \left|\psi_{\mathrm{o}}^{j}\right\rangle \}$, corresponding to the uncertain Hamiltonians, is the weighted average of the separate objective functions:
\begin{align}
J_\mathrm{f}\approx & \sum_{j}p_j\sum_{m}\left[ \left. \langle \psi ^{j}_\mathrm{o} \right|\left( I-i\tau \sigma^{(m)} _\mathrm{f}H^{(m)}_\mathrm{f} \right) U_{N}\cdots\left( I-i\tau \sigma^{(m)}_\mathrm{f}H^{(m)}_\mathrm{f} \right) U_2\left( I-i\tau \sigma^{(m)}_\mathrm{f}H^{(m)}_\mathrm{f} \right) U_1\left| \psi _{\mathrm{i}}^{j} \right. \rangle \times h.c. \right] 
\notag\\
&-\sum_{j}p_j\sum_{m}( \langle \psi_\mathrm{o}^{j} |U_{1\rightarrow N}\left| \psi_{\mathrm{i}}^{j}\right. \rangle \times h.c.)
\notag\\
& +\sum_{j}p_j\sum_{m}\left( i\tau \sigma^{(m)}_\mathrm{f}\sum_{i=1}^{N}{\left. \langle \psi^{j} _\mathrm{o} \right|U_{i+1\rightarrow N}H^{(m)}_\mathrm{f}U_{1\rightarrow i}\left| \psi _{\mathrm{i}}^{j} \right. \rangle \left. \langle \psi _{\mathrm{i}}^{j} \right|U_{1\rightarrow N}^{\dagger}\left| \psi^{j} _\mathrm{o} \right. \rangle}+h.c. \right). 
\end{align}
Here, $p_j$ is the weight of the $j$-th constraint in the optimization.

\subsubsection{Decoherence  noise}
Here, we further present the derivation of the fidelity loss induced by decoherence noise and we also incorporate this influence into the objective function. Iteratively, the closed-form expression of $\rho_1(T)$ in Equ.~\ref{PerturbationExpansion} can be obtained as:
\begin{align}
    &\rho _1\left( T \right) 
\notag\\
=&\rho _1\left( 0 \right) +\int_0^T{dt\mathscr{L} \left( \rho _0\left( t \right) \right)}-i\int_0^T{d\tau _1\left[ \widetilde{H}\left( \tau _1 \right) ,\rho _1\left( \tau _1 \right) \right]}
\notag\\
=&\int_0^T{dt\mathscr{L} \left( \rho _0\left( t \right) \right)}-i\int_0^T{d\tau _1}\int_0^{\tau _1}{dt}\left[ \widetilde{H}\left( \tau _1 \right) ,\mathscr{L} \left( \rho _0\left( t \right) \right) \right] +\cdots
\notag\\
&+\left( -i \right) ^{\boldsymbol{n}}\int_0^T{d\tau _1}\int_0^{\tau _1}{d\tau _2}\cdots\int_0^{\tau _{n-1}}{d\tau _n}\int_0^{\tau _n}{dt}\left[ \widetilde{H}\left( \tau _1 \right) ,\left[ \widetilde{H}\left( \tau _2 \right) ,\left[ \cdots\left[ \widetilde{H}\left( \tau _n \right) ,\mathscr{L} \left( \rho _0\left( t \right) \right) \right] \cdots \right] \right] \right] +\cdots
\notag\\
=&\int_0^Tdt\Big\{\mathscr{L} \left( \rho _0\left( t \right) \right)-i\int_t^T{d\tau _1}\left[ \widetilde{H}\left( \tau _1 \right) ,\mathscr{L} \left( \rho _0\left( t \right) \right) \right] +\cdots
\notag\\
&+\left( -i \right) ^{\boldsymbol{n}}\int_t^T{d\tau _1}\int_t^{\tau _1}{d\tau _2}\cdots\int_t^{\tau _{n-1}}{d\tau _n}\left[ \widetilde{H}\left( \tau _1 \right) ,\left[ \widetilde{H}\left( \tau _2 \right) ,\left[ \cdots\left[ \widetilde{H}\left( \tau _n \right) ,\mathscr{L} \left( \rho _0\left( t \right) \right) \right] \cdots \right] \right] \right] +\cdots\Big\}.
\end{align}
Here, $\rho_1(0)=0$. Similar to Equ.~\ref{Dyson},  the evolution within the curly brackets corresponds to the state $\mathscr{L} \left( \rho _0\left( t \right) \right)$ evolving from the initial time $t$ to time $T$, followed by the integration over different initial time $t$. The final state can be written in a discrete form as:
\begin{align}
&\rho _1\left( T \right) 
\notag\\
\approx & \tau\sum_{j=1}^{N}{\widetilde{U}_{j+1\rightarrow N}}\mathscr{L} \left( \rho _{0,j} \right) \widetilde{U}_{j+1\rightarrow N}^{\dagger}
\notag\\
=&\tau\sum_{j=1}^{N}{\widetilde{U}_{j+1\rightarrow N}}\mathscr{L} \left( \widetilde{U}_{0\rightarrow j}\rho _{0,0}\widetilde{U}_{0\rightarrow j}^{\dagger} \right) \widetilde{U}_{j+1\rightarrow N}^{\dagger}
\notag\\
\approx & \tau\sum_{j=1}^{N}{U_{j+1\rightarrow N}}\mathscr{L} \left( U_{0\rightarrow j}\rho _{0,0}U_{0\rightarrow j}^{\dagger} \right) U_{j+1\rightarrow N}^{\dagger}.
\end{align}
Since both decoherence noise and parameter fluctuations are weak, higher-order terms that include both types of noise have been neglected. To summarize, the objective function related to the decoherence noise is 
\begin{align}\label{inf_Lindblad_1}
 &J_\mathrm{d}
\notag\\
=& tr(\left| \psi_\mathrm{o} \right. \rangle \left. \langle \psi_\mathrm{o} \right|\rho_1(T))
\notag\\
\approx&\tau\sum_{j=1}^{N}{\langle\psi_\mathrm{o}| U_{j+1\rightarrow N}}\mathscr{L} \left( U_{0\rightarrow j}\rho _{0,0}U_{0\rightarrow j}^{\dagger} \right) U_{j+1\rightarrow N}^{\dagger}| \psi_\mathrm{o} \rangle.
\end{align}
To provide a more intuitive representation for the subsequent algorithmic calculations, Equ.~\ref{inf_Lindblad_1} can be further expressed in the following form under weak noise approximation:
\begin{align}\label{sequ:Phi_d_1}
& J_\mathrm{d}
\notag\\
\approx& \sum_{m=1}^v{\sum_{j=1}^{N}{\langle \psi_\mathrm{o}|U_{j+1\rightarrow N}}}\sqrt{\kappa _m\tau}L_mU_{0\rightarrow j}\rho _{0,0}U_{0\rightarrow j}^{\dagger}L_{m}^{\dagger}\sqrt{\kappa _m\tau}U_{j+1\rightarrow N}^{\dagger}|\psi_\mathrm{o}\rangle 
-\langle \psi_\mathrm{o}|U_{0\rightarrow N}\rho _{0,0}U_{0\rightarrow N}^{\dagger}|\psi_\mathrm{o}\rangle 
\notag\\
&+\sum_{m=1}^v{\sum_{j=1}^{N}{\langle \psi_\mathrm{o}|U_{j+1\rightarrow N}}}\left( \frac{1}{\sqrt{vN}}I-\frac{\sqrt{vN}}{2}\kappa _m\tau L_{m}^{\dagger}L_m \right) U_{0\rightarrow j}\rho _{0,0}U_{0\rightarrow j}^{\dagger}\left( \frac{1}{\sqrt{vN}}I-\frac{\sqrt{vN}}{2}\kappa _m\tau L_{m}^{\dagger}L_m \right) U_{j+1\rightarrow N}^{\dagger}|\psi_\mathrm{o}\rangle 
\notag\\
\approx& \sum_{m=1}^v{\sum_{j=1}^{N}{(\langle \psi_\mathrm{o}|U_{j+1\rightarrow N}}}\sqrt{\kappa _m\tau}L_mU_{0\rightarrow j}| \psi _{\mathrm{i}}\rangle\times c.c.)
-(\langle \psi_\mathrm{o}|U_{0\rightarrow N}| \psi _{\mathrm{i}}\rangle\times c.c.)
\notag\\
&+\sum_{m=1}^v{\sum_{j=1}^{N}{[\langle \psi_\mathrm{o}|U_{j+1\rightarrow N}}}\left( \frac{1}{\sqrt{vN}}I-\frac{\sqrt{vN}}{2}\kappa _m\tau L_{m}^{\dagger}L_m \right) U_{0\rightarrow j}| \psi _{\mathrm{i}}\rangle\times c.c.].
\end{align}
Here, $v$ is the number of independent decoherence noise sources and $c.c.$ refers to the complex conjugate of the expression within the parentheses or the square brackets. Equation~\ref{sequ:Phi_d_1} can be understood intuitively by representing the Lindblad master equation with the Kraus operators~\cite{jacobs2014quantum}. The first term is the summation of the states that suffer a single jump $\sqrt{\kappa _m\tau}L_m$ at different times. The second and the third terms are the first-order terms in the backaction evolution at different times. The ``no jump operator" here is calculated as $\left( \frac{1}{\sqrt{vN}}I-\frac{\sqrt{vN}}{2}\kappa _m\tau L_{m}^{\dagger}L_m \right)$ instead of $\left( I-\frac{1}{2}\kappa _m\tau L_{m}^{\dagger}L_m \right)$ to improve the accuracy of the numerical calculation. Written in this form, the objective function can be calculated with the trajectory instead of the density matrix, and the detail will be shown in the following section. To be reminded, the higher-order terms that indicate multiple jumps are neglected here since the decoherence is weak. 

To summarize, the fidelity of the final state $f_\mathrm{open}$ is equivalent to the objective function $J_\mathrm{open}$ under the weak noise approximation $J_\mathrm{open}\approx f_\mathrm{open}=\langle \mathrm{Tr}[\left| \psi_\mathrm{o} \right. \rangle \left. \langle \psi_\mathrm{o} \right|(\rho_0(T)+\rho_1(T))] \rangle$ and  
\begin{equation}
    J_\mathrm{open}=J_\mathrm{close}+J_\mathrm{d}+J_\mathrm{f}.
\end{equation}
Instead of maximizing the fidelity of the operation, the optimization can be implemented equivalently to minimize the infidelity $\widetilde{f}_\mathrm{open}=1-\sqrt{f_\mathrm{open}}$. In this situation, the objective function to be minimized in the Closed-GRAPE and approximate Open-GRAPE algorithm is:
\begin{equation}\label{equ:EquS18_ObjFunc}
    \Phi_\mathrm{close/open}=1-\sqrt{J_\mathrm{close/open}}.
\end{equation}

\subsection{The Algorithm to Calculate the Objective Function}
Similar to the Closed-GRAPE algorithm in Ref.~\cite{khaneja2005optimal}, the key idea of  the approximate Open-GRAPE algorithm is to calculate the objective function 
\begin{equation}
    \Phi_\mathrm{open}=1-\sqrt{J_\mathrm{close}+J_\mathrm{d}+J_\mathrm{f}}
\end{equation}
(related to Equs.~\ref{inf_0th}, \ref{inf_fluc_2}, and \ref{inf_Lindblad_1}) and the corresponding gradients
\begin{align}
    \frac{\partial \Phi_\mathrm{open}}{\partial u_{c,j}^{(k)}}= -\frac{1}{2}\frac{1}{1-\Phi_\mathrm{open}}(\frac{\partial J_\mathrm{close}}{\partial u_{c,j}^{(k)}}+\frac{\partial J_\mathrm{d}}{\partial u_{c,j}^{(k)}}+\frac{\partial J_\mathrm{f}}{\partial u_{c,j}^{(k)}}),
\end{align}
with $d\times1$-dimensional column vectors instead of $d\times d$-dimensional density matrices. In the following part, we will refer to these column vectors as ``states", even though they are not actual quantum states. The complete evolution process of these vectors will be referred to as ``trajectory". We will first show the calculation of the objective function with these trajectories in the following part.

\subsubsection{The algorithm for $J_\mathrm{close}$}
The calculation of $J_\mathrm{close}$ can be achieved through the calculation of the following trajectories:
\begin{enumerate}
\item $|A_{j}\rangle$ is the forward propagating state related to the evolution in the ideal closed system, where $\left|A_{0}\right\rangle =\left|\psi_{\mathrm{i}}\right\rangle $ and $\left|A_{j}\right\rangle =U_{j}\left|A_{j-1}\right\rangle $ ($1\leq j\leq N$);
\item  $\left|Z_{j}\right\rangle $ is the backward propagating state related to the evolution  in the ideal closed system, where $\left|Z_{N}\right\rangle =\left|\psi_\mathrm{o}\right\rangle $ and $\left|Z_{N-j}\right\rangle =U_{N-j+1}^{\dagger}\left|Z_{N-j+1}\right\rangle $ ($1\leq j\leq N$).
\end{enumerate}
The calculations of these two trajectories are the same in both the approximate Open-GRAPE algorithm and the Closed-GRAPE algorithm. The objective function of the Closed-GRAPE algorithm $J_\mathrm{close}$ in Equ. \ref{inf_0th} can be calculated as:
\begin{align}\label{sequ:Phi_close}
J_\mathrm{close}= \langle Z_{j} \lvert A_{j}\rangle \langle A_{j} \lvert Z_{j}\rangle
\end{align}
for arbitrary $j$ satisfying $1\leq j\leq N$.

\subsubsection{The algorithm for $J_\mathrm{f}$}
The calculation of the uncertain parameter parts in the objective function, i.e., $J_\mathrm{f}$  in  Equ.~\ref{inf_fluc_2}, can be achieved through the following trajectories:
\begin{enumerate}
    \item $\left|B_{j}^{m}\right\rangle $ is the forward propagating state related to the uncertainty of $H^{(m)}_\mathrm{f}$, where $\left|B_{0}^{m}\right\rangle =\left|\psi_{\mathrm{i}}\right\rangle$ and $\left|B_{j}^{m}\right\rangle =\left(I-i\tau\sigma_\mathrm{f}^{(m)}H^{(m)}_\mathrm{f}\right)U_{j}\left|B_{j-1}^{m}\right\rangle $ ($1\leq j\leq N$);
    \item  $\left|Y_{j}^{m}\right\rangle $ is the backward propagating state related to the uncertainty of $H_\mathrm{f}^{(m)}$, where $\left|Y_{N}^{m}\right\rangle =\left|\psi_\mathrm{o}\right\rangle $ and $\left|Y_{N-j}^{m}\right\rangle =U_{N-j+1}^{\dagger}\left(I+i\tau\sigma^{(m)}_\mathrm{f}H^{(m)}_\mathrm{f}\right)\left|Y_{N-j+1}^{m}\right\rangle $ ($1\leq j\leq N$);
    \item $\left|C_{j}\right\rangle $ is another forward propagating state related to the first-order uncertainty of $H_\mathrm{f}^{(m)}$, where $\left|C_{0}\right\rangle =\left|\psi_{\mathrm{i}}\right\rangle $, $\left|C_{1}\right\rangle =\sum_{m}\left(-i\tau\sigma^{(m)}_\mathrm{f}H^{(m)}_\mathrm{f}\right)\left|A_{1}\right\rangle $, and $\left|C_{j}\right\rangle =U_{j}\left|C_{j-1}\right\rangle +\sum_{m}\left(-i\tau\sigma^{(m)}_\mathrm{f}H^{(m)}_\mathrm{f}\right)\left|A_{j}\right\rangle $
($2\leq j\leq N$); 
\item $\left|X_{j}\right\rangle $ is another backward propagating state related to the first-order uncertainty of $H^{(m)}_\mathrm{f}$, where $\left|X_{N}\right\rangle =\sum_{m}\left(i\tau\sigma^{(m)}_\mathrm{f}H^{(m)}_\mathrm{f}\right)\left|\psi_\mathrm{o}\right\rangle $, $\left|X_{N-j}\right\rangle =U_{N-j+1}^{\dagger}\left|X_{N-j+1}\right\rangle +\sum_{m}\left(i\tau\sigma^{(m)}_\mathrm{f}H^{(m)}_\mathrm{f}\right)\left|Z_{N-j}\right\rangle $ ($1\leq j\leq N-1$), and $\left|X_{0}\right\rangle =U_{1}^{\dagger}\left|X_{1}\right\rangle $.
\end{enumerate}

With the trajectories above, the term $J_\mathrm{f}$  in Equ. \ref{inf_fluc_2} can also be calculated as:
\begin{align}\label{sequ:Phi_J}
J_\mathrm{f}
=& \sum_{m}\left(\langle Y_{j}^{m} \lvert B_{j}^{m}\rangle \langle B_{j}^{m}\lvert Y_{j}^{m}\rangle -\left\langle Z_{j}\lvert A_{j}\right\rangle \left\langle A_{j}\lvert Z_{j}\right\rangle \right)
\notag\\
& -\left[\left\langle Z_{j}\lvert C_{j}\right\rangle \left\langle A_{j}\lvert Z_{j}\right\rangle 
+\left\langle X_{j} \lvert A_{j}\right\rangle \left\langle A_{j} \lvert Z_{j}\right\rangle 
-\left\langle Z_{j}\right|\sum_{m}\left(-i\tau\sigma^{(m)}_\mathrm{f}H^{(m)}_\mathrm{f}\right)\left|A_{j}\right\rangle \left\langle A_{j} \lvert Z_{j}\right\rangle +c.c. \right],
\end{align}
with arbitrary $j$ satisfying $1\leq j\leq N$. 

\subsubsection{The algorithm for $J_\mathrm{d}$}
The calculation of the decoherence part in the objective function, i.e., $J_\mathrm{d}$ in Equ. \ref{inf_Lindblad_1}, can be achieved through the following trajectories:
\begin{enumerate}
    \item $\left|D_{j}\right\rangle $ is the forward propagating state related to all decoherence noises, where $\left|D_{0}\right\rangle =\left|\psi_{\mathrm{i}}\right\rangle $, $\left|D_{1}\right\rangle =\left(l_{1}+\zeta_{1}\right)\left|A_{1}\right\rangle $, and $\left|D_{j}\right\rangle =U_{j}\left|D_{j-1}\right\rangle +\left(l_{j}+\zeta_{j}\right)\left|A_{j}\right\rangle $ ($2\leq j\leq N$) ;
\item  $\left|W_{j}\right\rangle $ is the backward propagating state related to all decoherence noises, where $\left|W_{N}\right\rangle =\left(l_{N}^{\dagger}+\zeta_{N}^{\dagger}\right)\left|\psi_\mathrm{o}\right\rangle $, $\left|W_{N-j}\right\rangle =U_{N-j+1}^{\dagger}\left|W_{N-j+1}\right\rangle +\left(l_{N-j}^{\dagger}+\zeta_{N-j}^{\dagger}\right)\left|Z_{N-j}\right\rangle $ ($1\leq j\leq N-1$), and $\left|W_{0}\right\rangle =U_{1}^{\dagger}\left|W_{1}\right\rangle $.
\end{enumerate}
Here, $l_{j}=\sum_{m}\kappa_m\tau L_{m}$$\left\langle A_{j}\right|L_{m}^{\dagger}\left|Z_{j}\right\rangle $,  $\zeta_{j}=\sum_{m}\xi_{m}\left\langle A_{j}\right|\xi_{m}^{\dagger}\left|Z_{j}\right\rangle $, $\xi_{m}=\frac{1}{\sqrt{vN}}I-\frac{\sqrt{vN}}{2}\kappa_m \tau L_{m}^{\dagger}L_{m}$, and $v$ is the number of the independent decoherence noise sources.
With these trajectories, $J_\mathrm{d}$ in Equ. \ref{inf_Lindblad_1} can also be calculated as:
\begin{align}\label{sequ:Phi_d}
J_\mathrm{d}  
=\left\langle Z_{j} \lvert D_{j}\right\rangle +\left\langle W_{j}\lvert A_{j}\right\rangle -\left\langle Z_{j}\right|\left(l_{j}+\zeta_{j}\right)\left|A_{j}\right\rangle -\left\langle Z_{j} \lvert A_{j}\right\rangle \left\langle A_{j} \lvert Z_{j}\right\rangle,
\end{align}
for arbitrary $j$ satisfying $1\leq j\leq N$.

To be reminded, the calculation of $\left|A_{j}\right\rangle ,\left|Z_{j}\right\rangle ,|B_{j}^{m}\rangle$, and $|Y_{j}^{m}\rangle $ can be evaluated in parallel, and so can $\left|C_{j}\right\rangle ,\left|X_{j}\right\rangle ,\left|D_{j}\right\rangle $, and $\left|W_{j}\right\rangle $.

\subsection{The Gradient of the Objective Function}
Calculating the gradient efficiently is essential for the GRAPE algorithm. In this section, we will show the gradient corresponding to the objective function, i.e., the influence induced by the uncertain parameters in Equ. \ref{inf_fluc_2} and decoherence noise in Equ.~\ref{inf_Lindblad_1}.

\subsubsection{The gradient of $J_\mathrm{close}$ and the algorithm}
For small $\tau$ satisfying $\tau\ll||H_{0}+\sum_{k}u^{(k)}_{c,j}H_{c}||^{-1}$, we have $\delta U_{j}=-i\tau \delta u^{(k)}_{c,j} H^{(k)}_{c}  U_{j}$. The approximate gradient in the Closed-GRAPE algorithm~\cite{khaneja2005optimal} can be calculated as:
\begin{align}\label{grad_0th}
    \frac{\delta J_\mathrm{close}}{\delta u^{(k)}_{c,j}}=-i\tau \left. \langle \psi_\mathrm{o} \right|U_{j+1\rightarrow N}H^{(k)}_cU_{1\rightarrow j}\rho _0U_{1\rightarrow N}^{\dagger}\left| \psi_\mathrm{o} \right. \rangle +c.c.
\end{align}
In the Closed-GRAPE algorithm, the gradient can be calculated through the forward propagation of the initial state and the backward propagation of the target state, i.e., $\{|A_{j}\rangle\}$ and $\{|Z_{j}\rangle\}$ with $j=1, 2,\cdots, N$. Since $\partial \left|A_j\right\rangle/ \partial u^{(k)}_{c,j}=-i\tau H_c^{(k)} \left|A_j\right\rangle$ and $\partial \left|Z_j\right\rangle/ \partial u^{(k)}_{c,j}=0$, this gradient can also be calculated according to Equ.~\ref{sequ:Phi_close} as:
\begin{align}\label{algorithm_0th}
    \frac{\delta J_\mathrm{close}}{\delta u^{(k)}_{c,j}}=-i\tau \langle Z _{j} | H^{(k)}_c |A_{j}\rangle \langle A_{j}| Z_{j} \rangle +c.c.
\end{align}
\subsubsection{The gradient of $J_\mathrm{f}$ and the algorithm}
The gradient related to the uncertain parameters can be calculated similarly according to Equ.~\ref{sequ:Phi_J} as:
\begin{align}\label{algorithm_fluc}
\frac{\delta J_\mathrm{f}}{\delta u^{(k)}_{c,j}}
=&\sum_m{\left( -i\tau \left. \langle Y_{j}^{m} \right|H^{(k)}_c\left| B_{j}^{m} \right. \rangle \left. \langle B_{j}^{m}|Y_{j}^{m} \right. \rangle +i\tau \left. \langle Z_j \right|H^{(k)}_c\left| A_j \right. \rangle \left. \langle A_j|Z_j \right. \rangle +c.c. \right)}
\notag\\
&+[i\tau \left. \langle Z_j \right|H^{(k)}_c\left| C_j \right. \rangle \left. \langle A_j|Z_j \right. \rangle -i\tau \left. \left. \langle Z_j|C_j \right. \rangle \langle A_j \right|H^{(k)}_c\left| Z_j \right. \rangle +c.c.]
\notag\\
&+[i\tau \left. \langle X_j \right|H^{(k)}_c\left| A_j \right. \rangle \left. \langle A_j|Z_j \right. \rangle -i\tau \left. \left. \langle X_j|A_j \right. \rangle \langle A_j \right|H^{(k)}_c\left| Z_j \right. \rangle +c.c.]
\notag\\
&+[-i\tau \left. \langle Z_j \right|\sum_m{\left( -i\tau \sigma^{(m)}_\mathrm{f}H^{(m)}_\mathrm{f} \right)}H^{(k)}_c\left| A_j \right. \rangle \left. \langle A_j|Z_j \right. \rangle +i\tau \left. \langle Z_j \right|\sum_m{\left( -i\tau \sigma^{(m)}_\mathrm{f}H^{(m)}_\mathrm{f} \right)}\left| A_j \right. \rangle \left. \langle A_j \right|H^{(k)}_c\left| Z_j \right. \rangle +c.c.].
\end{align}
The derivation here relies on the following equations: $\partial \left|B^m_j\right\rangle/ \partial u^{(k)}_{c,j}=-i\tau H_c^{(k)} \left|B^m_j\right\rangle$, $\partial \left|Y^m_j\right\rangle/ \partial u^{(k)}_{c,j}=0$, $\partial \left|C_j\right\rangle/ \partial u^{(k)}_{c,j}=-i\tau H_c^{(k)} \left|C_j\right\rangle$, and $\partial \left|X_j\right\rangle/ \partial u^{(k)}_{c,j}=0$.

 \subsubsection{The gradient of $J_\mathrm{d}$ and the algorithm}
The gradient corresponding to the decoherence noise is 
 \begin{align}\label{grad_Lindblad}
\frac{\delta J_\mathrm{d}}{\delta u^{(k)}_{c,j}}=-i\tau ^2\sum_{m=j}^{N}{\langle \psi_\mathrm{o}|U_{m+1\rightarrow N}}\mathscr{L} \left( U_{j+1\rightarrow m}H^{(k)}_cU_{0\rightarrow j}\rho _{0,0}U_{0\rightarrow m}^{\dagger} \right) U_{m+1\rightarrow N}^{\dagger}|\psi_\mathrm{o}\rangle 
\notag\\
-i\tau ^2\sum_{m=1}^{j-1}{\langle \psi_\mathrm{o}|U_{j+1\rightarrow N}}H^{(k)}_cU_{m+1\rightarrow j}\mathscr{L} \left( U_{0\rightarrow m}\rho _{0,0}U_{0\rightarrow m}^{\dagger} \right) U_{m+1\rightarrow N}^{\dagger}|\psi_\mathrm{o}\rangle +h.c.
 \end{align}

Similar to Equ.~\ref{sequ:Phi_d_1}, the gradient related to the decoherence noise can be calculated as:

\begin{align}\label{algorithm_Lindblad}
\frac{\delta J_\mathrm{d}}{\delta u^{(k)}_{c,j}}=&-\left\langle Z_{j}\right|i\tau H^{(k)}_{c}\left|D_{j}\right\rangle -\left\langle W_{j}\right|i\tau H^{(k)}_{c}\left|A_{j}\right\rangle +\left\langle Z_{j}\right|\left(l_{j}+\zeta_{j}\right)i\tau H^{(k)}_{c}\left|A_{j}\right\rangle
\notag\\
&+\langle Z_j |i\tau H^{(k)}_c| A_j \rangle \langle A_j | Z_j \rangle + c.c.
\end{align}
The complete gradient can be obtained through the summation of the three gradients together in Equs.~\ref{algorithm_0th}, \ref{algorithm_fluc}, and \ref{algorithm_Lindblad}, while the gradient for multiple sets of initial and target states can be obtained through the weighted summation of the individual complete gradients.
Similar to the Closed-GRAPE algorithm, the update of the control parameters is realized through a combination of the L-BFGS-B algorithm~\cite{byrd1995limited} and the gradient in the approximate Open-GRAPE algorithm.

\subsection{Complexity Analysis}
To better illustrate the computational efficiency of the approximate Open-GRAPE algorithm, we show the computational complexity of calculating the gradient during each iteration in the optimization. The results in the following sections indicate that the complexity of the approximate Open-GRAPE algorithm exhibits only a modest increase compared to the Closed-GRAPE algorithm. As is common in algorithmic analysis,  computational complexity is primarily determined by the number of multiplications involved in matrix operations. Specifically, the computational complexity is $\mathcal{O}(abc)$ for the multiplication of an $a\times b$-dimensional matrix and a $b\times c$-dimensional matrix. 

\subsubsection{Matrix-vector exponential method}
In the Closed-GRAPE algorithm, the objective function in Equ.~\ref{sequ:Phi_close} and the gradient in Equ.~\ref{grad_0th} can be calculated through the forward propagating trajectories of the initial states and the backward propagating trajectories of the target states, i.e., $\{\left|A_j\right\rangle= U_{1 \rightarrow j}\left|\psi_{\mathrm{i}}\right\rangle \}$ and $\{\left|Z_j\right\rangle = U_{j+1 \rightarrow N}^{\dagger} \left|\psi_{\mathrm{o}}\right\rangle \}$. 
This calculation can be divided into two parts. The first part is to calculate the matrix exponential of the unitary operators $U=\textrm{exp}\left(-i\tau H\right)$, from the Hamiltonian $H=H_{0}+\sum_{k}u^{(k)}_{c}H^{(k)}_{c}$ with the control parameters $\{ u^{(k)}_{c} \} $. The second part is to calculate the propagations of the initial and target states with these operators.

For a $d$-dimensional system with the control pulses equally discretized into $N$ steps, the computational complexity of the matrix exponential is $\mathcal{O}(Nd^{3})$ in each iteration. Since the propagating trajectories of the initial state and target state can be represented with  $d\times 1$-dimensional vectors, the complexity of calculating these two propagating trajectories is only $\mathcal{O}(2Nd^2)$ in each iteration. Therefore, the exponential of the Hamiltonian matrix will become the most time-consuming part, and this will limit the application in high dimensional systems with large $d$. Fortunately, this can be improved with the matrix-vector exponential method shown in Ref.~\cite{abdelhafez2019gradient} and the earlier references~\cite{kuprov2007polynomially,kuprov2008polynomially}. The key idea is to calculate the unitary evolution with the Taylor expansion of the unitary operator when $\tau$ satisfying $\tau \ll ||H||^{-1}$. Specifically, the calculation of the state $U\left|\psi\right\rangle=\left[I+(-i\tau )H+\frac{(-i\tau )^2}{2!}H^2+\frac{(-i\tau )^3}{3!}H^3+\cdots\right]\left|\psi\right\rangle$ can be realized through the iterative calculation of  $\left|\phi_0\right\rangle=\left|\psi\right\rangle$ and $\left|\phi_m\right\rangle=\frac{-i\tau}{m}H\left|\phi_{m-1}\right\rangle$ for $m>1$. This approximation can be truncated at the $n_\mathrm{Taylor}$-th order of the unitary operator, and the result becomes $U\left|\psi\right\rangle=\sum_{m=0}^{n_\mathrm{Taylor}} \left|\phi_m\right\rangle+\mathcal{O}\left(\frac{\left|\tau H\right|^{n_\mathrm{Taylor}}}{(n_\mathrm{Taylor}+1)!} \right)$. Under this approximation, the complexity of the unitary evolution, i.e., $\sum_{m=0}^{n_\mathrm{Taylor}} \left|\phi_m\right\rangle$, is only $\mathcal{O}(n_\mathrm{Taylor} d^2)$ with $n_\mathrm{Taylor} \ll d$ in most applications. To mitigate the accumulation of numerical calculation errors, the state can be normalized after each computation. The computational complexity of the normalization is only $\mathcal{O}(d)$, which can be neglected.

\begin{table}[h]
    \centering
    \begin{tabular}{|c|c|c|c|c|}
         \hline
         state vector &
         $|A_{j}\rangle$ or $|Z_{j}\rangle$ &
         $|B_{j}^{m}\rangle$ or $|Y_{j}^{m}\rangle$&
         $|C_{j}\rangle $ or $|X_{j}\rangle$ &
         $|D_{j}\rangle $ or $|W_{j}\rangle$ \\
         \hline
         complexity &
         $\mathcal{O}(n_\mathrm{Taylor} d^{2})$&
         $\mathcal{O}[ (n_\mathrm{Taylor}+1) d^2]$&
         $\mathcal{O}[  (n_\mathrm{Taylor}+1)  d^2]$&
         $\mathcal{O}[(2v+1+n_\mathrm{Taylor})  d^2]$\\
         \hline
    \end{tabular}
    \caption{The computational complexity of a single state vector in the calculation of the propagation.}
    \label{tab:complexity}
\end{table}

\subsubsection{The complexity of the Closed-GRAPE algorithm}
With the matrix-vector exponential method mentioned above, the calculation of trajectories $\{\left|A_j\right\rangle\}$ and $\{\left|Z_j\right\rangle\}$ is $\mathcal{O}(2 n_\mathrm{Taylor} N d^2)$, the calculation of the objective function in Equ.~\ref{sequ:Phi_close} is $\mathcal{O}(2d)$, and the calculation of the gradient in Equ.~\ref{algorithm_0th} is $\mathcal{O}(n_\mathrm{control} N d^2)$. Here, $n_{\mathrm{control}}$ is the number of the control Hamiltonians to be optimized. Therefore, the total complexity is $\mathcal{O}\left[(2 n_\mathrm{Taylor}+n_\mathrm{control})Nd^2\right]$ for the Closed-GRAPE algorithm.

\subsubsection{The complexity of the approximate Open-GRAPE algorithm}
Although the calculation is more complicated in the approximate Open-GRAPE algorithm than in the Closed-GRAPE algorithm, there is only a modest linear increase in the complexity.  The complexity of calculating a single state vector is shown in the table~\ref{tab:complexity}, and the complete propagation includes $N$ states for trajectories  $\{\left|A_j\right\rangle\}$, $\{\left|Z_j\right\rangle\}$, $\{\left|C_{j}\right\rangle \}$, $\{\left|X_{j}\right\rangle \}$, $\{\left|D_{j}\right\rangle \}$,  $\{\left|W_{j}\right\rangle \}$,  $\{\left|B_{j}^{m}\right\rangle \}$, and $\{\left|Y_{j}^{m}\right\rangle \}$, with $1\leq m \leq f$. Here, $f$ is the number of uncertain Hamiltonians. To calculate the objective function, $J_\mathrm{close}$, $J_\mathrm{d}$, and $J_\mathrm{f}$ can be calculated through Equs.~\ref{algorithm_0th}, \ref{algorithm_fluc}, and \ref{algorithm_Lindblad}, and the corresponding complexities are $\mathcal{O}(2d)$, $\mathcal{O}(d^2)$, and $\mathcal{O}(2 v d^2)$, respectively. These three terms can be neglected compared with the other terms during the calculation of the gradients. The gradients can be calculated through Equs.~\ref{grad_0th}, \ref{algorithm_fluc}, and \ref{grad_Lindblad} with the corresponding complexities being $\mathcal{O}(n_{\mathrm{control}} N d^2)$,  $\mathcal{O}[(2f+8)n_{\mathrm{control}} N d^2]$, and $\mathcal{O}[(2v+4)n_{\mathrm{control}} N d^2]$, respectively. Therefore, the total complexity is $\mathcal{O}\{[(2v+2f+13)n_{\mathrm{control}}+2(f+3)n_\mathrm{Taylor}+2(f+2v+1)] N d^2\}$ for the approximate Open-GRAPE algorithm. In the main text, the numerical simulation is implemented with $v=2$, $f=2$, $n_{\mathrm{control}}=4$, and $n_\mathrm{Taylor}=20$. Consequently, the complexity is $\mathcal{O}(44 N d^2)$ and $\mathcal{O}(298 N d^2)$ for the Closed-GRAPE algorithm and the approximate Open-GRAPE algorithm, respectively. This implies that the calculation time is only around  $6.77$ times that of the Closed-GRAPE algorithm.

\subsection{Implementing Dynamical Decoupling with the approximate Open-GRAPE algorithm}

Quantum control techniques offer various methods to mitigate the negative effects of parameters uncertainties in quantum systems. One such class of techniques is dynamical decoupling. In this section, we present optimization results for a simple example. We demonstrate how the approximate Open-GRAPE algorithm achieves suppression of the detrimental impacts from parameter uncertainties in a manner similar to dynamical decoupling techniques~\cite{lidar2014review,barnes2022dynamically}. While the approximate Open-GRAPE approach does not provide an intuitive physical picture like dynamical decoupling, it offers greater versatility that it can be applied to a wide range of Hamiltonian uncertainties and also to the implementation of more complicated quantum operations.

We consider an example of the a qubit subject to a fluctuating transverse field described by the Hamiltonian
\begin{equation}
    \widetilde{H}(t)=u_c(t)H_c+\delta u_\mathrm{f}H_\mathrm{f},
\end{equation}
where $H_c=\frac{1}{2}\sigma_z$ is the control Hamiltonian representing a tunable energy splitting, and $H_\mathrm{f}=\hat{\sigma}_x$ is the uncertain Hamiltonian with $\langle\delta u_\mathrm{f}^2 \rangle=\sigma_\mathrm{f}^2$. The unitary evolution operator can be written as 
\begin{align}
    U(t)=\left(
    \begin{array}{cc}
         u_1(t) & -u^{*}_2(t) \\
         u_2(t) & u^*_1(t)
    \end{array}
    \right).
\end{align}
Here, $u_1(t)=\mathrm{e}^{-i\phi(t)/2}\left(g_0(t)-g_2(t)\delta u _\mathrm{f}^2\right)+\mathcal{O}(\delta u_\mathrm{f}^3)$ and $u_2(t)=-i\mathrm{e}^{i\phi(t)/2}g^{*}_1(t)\delta u _\mathrm{f}+\mathcal{O}(\delta u_\mathrm{f}^3)$, where $\phi(t)=\int_0^t u_c(\tau)d\tau$ is the rotation angle and $g_n(t)=\int_0^t \mathrm{e}^{i\phi(\tau)}g^{*}_{n-1}(\tau)d\tau$ are the error coefficients with $g_o(t)=1$. During the optimization, the unitary operator can be represented by several constrains of the form $\{\ket{\psi_{\mathrm{i}}^{j}}\mapsto\ket{\psi_{\mathrm{o}}^{j}}\}$, where $\ket{\psi_{\mathrm{i}}^{j}}=\alpha^j \left|g\right\rangle+\beta^j \left|e\right\rangle$ is the $j$-th initial state and  $\left|\Psi_\mathrm{o}^j\right\rangle=\alpha^j \mathrm{e}^{-i\phi/2} \left|g\right\rangle+\beta^j \mathrm{e}^{i\phi/2} \left|e\right\rangle$ is the corresponding target state with a target rotation angle $\phi=\phi(T)$. The fidelity of the final state is given by $|\left\langle\Psi_\mathrm{o}^j\right|U(T)\left|\Psi_\mathrm{i}^j\right\rangle|^2=1-2\mathrm{Re}[g_2(T)]\delta u _\mathrm{f}^2+4\left(\mathrm{Re}[-\alpha^{j*} \beta^j g_1(T)]\right)^2 \delta u _\mathrm{f}^2+\mathcal{O}(\delta u _\mathrm{f}^3)
$.

\begin{figure}
\begin{centering}
\includegraphics[scale=0.8]{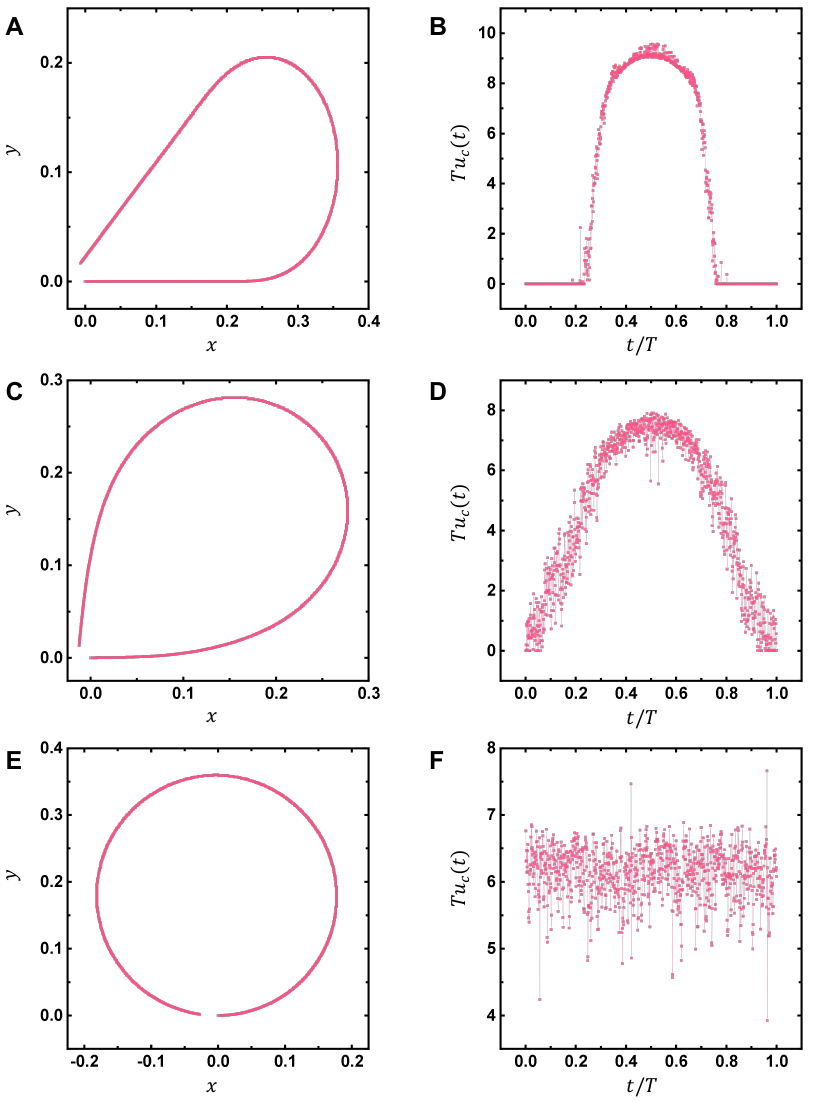}
\par\end{centering}
\caption{\label{fig:FigS1} \textbf{Geometric space curves and the corresponding pulses considering a fluctuating transverse field.} (\textbf{A}), (\textbf{C}), and (\textbf{E}) show the geometric space curves for target rotation angles $\phi=5\pi/4$, $3\pi/2$, and $2\pi$, respectively. (\textbf{B}), (\textbf{D}), and (\textbf{F}) show the optimized control pulses from the approximate Open-GRAPE algorithm corresponding to (A), (C), and (E), respectively.}
\end{figure}

As shown in previous works~\cite{zeng2018general,barnes2022dynamically}, when the error coefficient $g_1(T)=0$, the real part of $g_2(T)$ also vanishes, i.e., $\mathrm{Re}[g_2(T)]=0$. This condition eliminates the lowest-order effect of the uncertain parameter. To gain a geometric understanding of this condition, we can plot $g_1(t)$ in a Cartesian coordinates with $x=\mathrm{Re}[g_1(t)]$ and $y=\mathrm{Im}[g_1(t)]$. This curve is called the geometric space curve~\cite{barnes2022dynamically} and its shape is determined by the control pulses according to the definition above. A solution for $g_1(t)$ that eliminates the lowest-order effect corresponds to a closed curve that begins and ends at the origin. Interesting, the angle $\Delta \theta$ formed between the tangents at the start and end points of the curve is related to the required rotation angle $\phi$ by the equation $\phi=\Delta \theta+\pi$. By considering the lowest-order effects during the optimization, the approximate Open-GRAPE algorithm can also achieve comparable results. Figure~\ref{fig:FigS1} illustrates three examples of optimized control pulses and their corresponding curves for target rotation angles of $\phi=5\pi/4$, $3\pi/2$, and $2\pi$, respectively. As shown in the figure, these curves are nearly closed, indicating that the performance of the approximate Open-GRAPE algorithm against uncertain parameters is comparable to that of the dynamical decoupling techniques. To better compare with dynamical decoupling techniques, a penalty term is added during the optimization process to slightly smooth the control pulses. For more details on the penalty term (see the "Penalty term for pulse shape" section for more details). The optimization parameters are: $N_s=1000$, $\sigma_\mathrm{f}=1~\mathrm{MHz}$, the maximum allowable amplitude of the control Hamiltonian of $100~\mathrm{MHz}$, and duration times $T=0.18\pi~\mathrm{\mu s}$, $0.18\pi~\mathrm{\mu s}$ and $0.22\pi~\mathrm{\mu s}$ for $\phi=5\pi/4$, $3\pi/2$, and $2\pi$, respectively.

\section{Experimental Details}
\subsection{The Model of the Experimental Device}

As shown in Fig.2 of the main text, the experiment is implemented with an ancillary transmon qubit and one three-dimensional coaxial cavity. These components are selected from a larger system with multiple qubits and cavities as described in Ref.~\cite{cai2023protecting}. Throughout the experiment, no active control is performed on the remaining components and their dynamics can be ignored. The coaxial cavity with a long lifetime of $1300~\mu$s serves as the storage cavity for storing the logical states of the system. The transmon qubit serves as the ancillary mode with large energy gaps far detuned from the driving signal and can be regarded as a two-level qubit, providing necessary nonlinearity for universal control. The large metal pads of the transmon qubit is fabricated with tantalum films, exhibiting an energy relaxation time~\cite{wang2022towards} of $T_1=110~\mathrm{\mu}$s. The transmon qubit can be measured with high fidelity through its own readout cavity. In this system, both the storage and the readout cavities are dispersively coupled to the transmon qubit, and the Hamiltonian of the whole system can be written as:
\begin{equation}\label{ExpHamiltonian}
\frac{H}{\hbar}=-\frac{1}{2} \omega_\mathrm{q} \hat{\sigma}_z+\omega_\mathrm{s} \hat{a}_\mathrm{s}^{\dagger} \hat{a}_\mathrm{s}+\omega_\mathrm{r} \hat{a}_\mathrm{r}^{\dagger} \hat{a}_\mathrm{r}+\frac{1}{2} \chi_\mathrm{q r} \hat{a}_\mathrm{r}^{\dagger} \hat{a}_\mathrm{r} \hat{\sigma}_z+\frac{1}{2} \chi_\mathrm{q s} \hat{a}_\mathrm{s}^{\dagger} \hat{a}_\mathrm{s} \hat{\sigma}_z-\frac{K_\mathrm{s}}{2} \hat{a}_\mathrm{s}^{\dagger} \hat{a}_\mathrm{s}^{\dagger} \hat{a}_\mathrm{s} \hat{a}_\mathrm{s},
\end{equation}
where $\hat{a}_\mathrm{s}$ and $\hat{a}_\mathrm{r}$ are the bosonic annihilation operators of the storage cavity and the readout cavity, respectively; $\omega_\mathrm{q}$, $\omega_\mathrm{s}$, and $\omega_\mathrm{r}$ are the active mode frequencies of the transmon qubit, the storage cavity, and the readout cavity, respectively. Here, $\chi_\mathrm{qr}/2\pi=2.2~\mathrm{MHz}$ ($\chi_\mathrm{qs}/2\pi=1.00~\mathrm{MHz}$) is the dispersive coupling coefficient between the transmon qubit and the readout cavity mode (storage cavity mode), and $K_\mathrm{s}/2\pi=1.415$~kHz is the first-order self-Kerr coefficient of the storage cavity. We perform detailed characterization of these parameters, and the coherence property of each mode is also measured and shown in Table~\ref{Tab2}. 

\begin{table}[]
\begin{tabular}{|c|c|c|c|c|}
\hline
Mode & Frequency (GHz) & $T_1$ ($\mu$s) & $T_2$ ($\mu$s) & $n_\mathrm{th}$    \\ \hline
transmon qubit & 4.886        & 110& 130& 0.04 \\ \hline
storage cavity & 6.028          & 1300& --     & 0.03 \\ \hline
readout cavity & 7.576          & 0.86   & --     & --   \\ \hline
\end{tabular}
\caption{\label{Tab2}Parameters of the experimental device.}
\end{table}

Universal control of the storage cavity and the transmon qubit can be achieved with coherent drives of the transmon qubit and the storage cavity via the nonlinearity from their cross-Kerr interaction~\cite{krastanov2015universal,hu2019quantum}. The coherent drives are realized through external microwaves imported from transmission lines connected to the transmon qubit and the storage cavity~\cite{Blais2021}. For instance, a coherent drive of the storage cavity $H_\mathrm{d}/\hbar\propto iA(t)(a^{\dagger}e^{-i\omega_\mathrm{d} t-i\phi_\mathrm{d}}-a e^{i\omega_\mathrm{d} t+i\phi_\mathrm{d}})$ can be realized through a microwave drive with a frequency of $\omega_\mathrm{d}$, a phase of $\phi_\mathrm{d}$, and an amplitude of $A(t)$. Similar control can be extended to the transmon qubit. In total, the coherent drives can be realized with the following four independent control Hamiltonians:
\begin{align}
H^{(1)}_c= & \hat{\sigma}_x, \notag\\
H^{(2)}_c= & \hat{\sigma}_y, \notag\\
H^{(3)}_c= & a+a^{\dagger}, \notag\\
H^{(4)}_c= & i\left(a-a^{\dagger}\right),
\end{align}
where $\hat{\sigma}_x$ and $\hat{\sigma}_y$ are the Pauli-X and Pauli-Y operators of the transmon qubit, respectively. 

\subsection{The External Microwave Control System}
\begin{figure}
\begin{centering}
\includegraphics[scale=1.3]{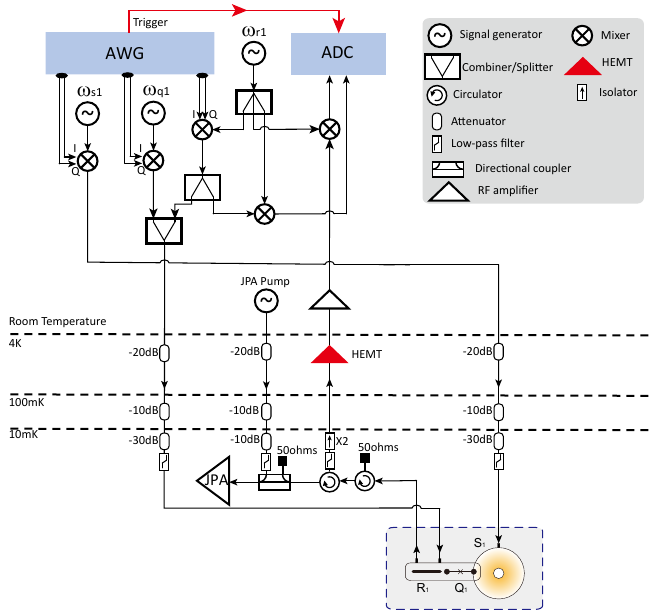}
\par\end{centering}
\caption{\label{fig:Fig8}Schematic of the external microwave control system of the experiment. }
\end{figure}

The schematic of the experimental system setup is shown in Fig.~\ref{fig:Fig8}. The experimental device is mounted in a dilution refrigerator with a base temperature lower than 10mK. The whole sample is enclosed by a shield made of high-$\mu$ metal to reduce the effect of the external magnetic field. We add low-pass filters and microwave attenuators along the input lines to further suppress the noise from higher-temperature plates and the external environment. In our experiment, the storage cavity, the readout cavity, and the qubit are all controlled by modulated signals, which are accomplished by an IQ mixer and a local microwave generator. The IQ signals at a few hundred $\mathrm{MHz}$ are generated by an arbitrary waveform generator (AWG).

A Josephson parametric amplifier (JPA) is utilized to enhance the readout performance. The JPA is connected to the output port of the readout cavity through two concatenated circulators that are used to avoid the reflected microwave signal. By carefully adjusting the bias current and the pump, the JPA is biased at a working point with a gain of about $20~\mathrm{dB}$ and a bandwidth of about $20~\mathrm{MHz}$, which assists in performing the fast high-fidelity single-shot readout of the qubit state. In addition, a high-electron-mobility transistor (HEMT) and other room-temperature amplifiers are used along the output line during the readout process. The output signal is then down-converted to
50 MHz, filtered with a bandpass filter, and finally digitized by an analog-to-digital converter (ADC).

\subsection{State Initialization and Gate Implementation}

\subsubsection{Cooling circuit}
\begin{figure}
\begin{centering}
\includegraphics[scale=0.5]{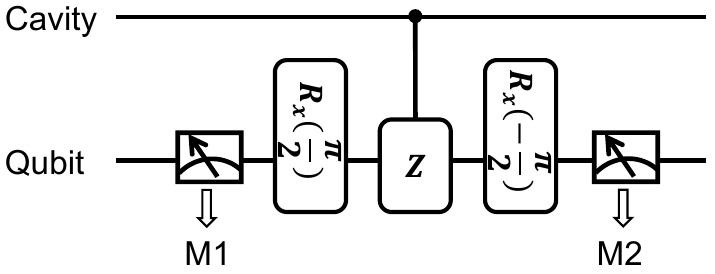}
\par\end{centering}
\caption{\label{fig:FigS2_Cooling}The cooling circuit in the initialization process. Two projective measurements, denoted as M1 and M2, are implemented on the qubit at the beginning and the end, respectively. The controlled-Z gate between the qubit and the cavity is realized through their dispersive coupling interaction with a duration of $\pi/\chi$.}
\end{figure}

As shown in Fig.~4 of the main text, the system should first be cooled down to the ground state $|g\rangle \otimes |0\rangle $ during the initialization process. This cooling process is achieved through system decay to the thermal equilibrium state, typically at least $15~\mathrm{ms}$ for each independent experiment. To mitigate the undesired thermal excitations, we implement the quantum circuit illustrated in Fig.~\ref{fig:FigS2_Cooling}. In this circuit, a quantum non-demolition (QND) measurement denoted as M1 is first performed on the transmon qubit to confirm its ground state $|g\rangle$, while the excited state $|e\rangle$ is removed through post-selection. 
Subsequently, a parity measurement of the storage cavity is implemented with a Ramsey-like pulse sequence [$R_x(\pi/2)$, CZ, $R_x(-\pi/2)$], where $R_x(\pm\pi/2)$ are unconditional qubit rotations of $\pm\pi/2$ around the $x$-axis and CZ is the controlled-Z gate between the qubit and the cavity with a duration of $\pi/\chi$. Finally, a second measurement M2 is performed to post-select the even parity state to ensure a vacuum state of the storage cavity. By removing the excitation populations based on the measurement results of both M1 and M2, the system is effectively cooled down to the desired ground state $|g\rangle \otimes |0\rangle $ with high fidelity. 


In this circuit, the $R_{x}(\pm\pi/2)$ gates applied to the transmon qubit are executed using Gaussian-shaped driving pulses with a duration of $20~\mathrm{ns}$. The shapes of these pulses are slightly adjusted using the ``Derivative Removal by Adiabatic Gate" technique~\cite{motzoi2009DRAG} to mitigate energy-level leakage. The necessary entanglement between the transmon and the storage cavity is realized by waiting for a period of $\pi/\chi=1~\mathrm{\mu}$s, where $\chi$ is the dispersive coupling strength $\chi_\mathrm{q s}$ in Equ.~\ref{ExpHamiltonian}. 

\subsubsection{Penalty term for pulse shape}
Both the Closed-GRAPE and approximate Open-GRAPE algorithms used in the Numerical Simulation section of the main text impose no additional restrictions on the pulse shape. However, in the practical experiment, limitations such as the bandwidth of the Arbitrary Waveform Generator (AWG) can result in distortion of the pulse shape. Therefore, some extra penalty terms should be added to the objective function to ensure the quality of the driving pulses from the AWG. In this superconducting platform, two extra penalty terms $P_a$ and $P_d$ are needed~\cite{ofek2016extending}, where
\begin{align}
    P_a = &\frac{A_a}{N}\times\sum_{j,k}\left( \exp{\left(\frac{u^{(k)}_{c,j}}{h_a}\right)^2}-1\right)
\end{align}
is to prevent the control amplitude from being too large and
\begin{align}
    P_d = &\frac{A_d}{N}\times\sum_{j,k}\left( \exp{\left(\frac{u^{(k)}_{c,j+1}-u^{(k)}_{c,j}}{h_d}\right)^2}-1\right)
\end{align}
is to punish large adjacent deviations, i.e., to reduce the high-frequency component in the driving pulse. Here, $N$ is the number of steps in the GRAPE algorithm, $A_a = A_d =0.1$ is the factor to determine the overall scaling of the penalty terms, $h_a$ is the threshold limiting the amplitude, and $h_d$ is the threshold limiting the different neighboring control terms. In the experiment, these values are chosen as $h_a=179~\mathrm{MHz}$ and  $h_d=22.4~\mathrm{MHz}$. To summarize, the experimental objective function is the summation of these two terms and the objective function in Equ.\ref{equ:EquS18_ObjFunc} , i.e., $\Phi_\mathrm{open}+P_a+P_d$. To be reminded, the gradient of these two terms can be calculated easily without calculating the propagation.

\subsection{State Tomography and Data Analysis}
To assess the performance of the Closed-GRAPE and approximate Open-GRAPE algorithms, we estimate the infidelity of the final state by conducting state tomography on the ancillary qubit.
When the density matrix $\rho$ of the ancillary qubit is reconstructed, the infidelity can be calculated as $\widetilde{f}=1-\sqrt{\left\langle \psi_\mathrm{o}\right|\rho \left| \psi_\mathrm{o} \right\rangle}$. Here, $\left|\psi_\mathrm{o}\right\rangle$ is the expected final state based on the repetition number $M$ of the $R_y(\pi)$  gate. For even values of $M$,  $\left|\psi_\mathrm{o}\right\rangle=(\left|g\right\rangle+\left|e\right\rangle)/\sqrt{2}$,  whereas for odd values of $M$,  $\left|\psi_\mathrm{o}\right\rangle=(\left|g\right\rangle-\left|e\right\rangle)/\sqrt{2}$.

We perform tomography on the ancillary qubit by executing pre-rotations followed by  $\hat{\sigma}_z$ projection measurements. To construct the density matrix
$\rho=\begin{pmatrix} \rho_{11}&\rho_{12}\\\rho_{21}&\rho_{22}\end{pmatrix}$,
four pre-rotation gates are implemented including $I$, $R_x(\pi)$, $R_x(\pi/2)$, $R_{y}(\pi/2)$. Here, the first two gates provide information about the diagonal elements $(\rho_{11}$ and $\rho_{22})$ of the density matrix, while the latter two gates provide information about the off-diagonal elements $(\rho_{21}$ and $\rho_{12})$. To ensure measurement accuracy, each pre-rotation gate is implemented with $5000$ repetitions. Based on these results, we can reconstruct the density matrix using maximum likelihood estimation to ensure that the estimated density matrix is Hermitian, positive semi-definite, and has a trace of 1.

\begin{figure}
\begin{centering}
\includegraphics[scale=0.75]{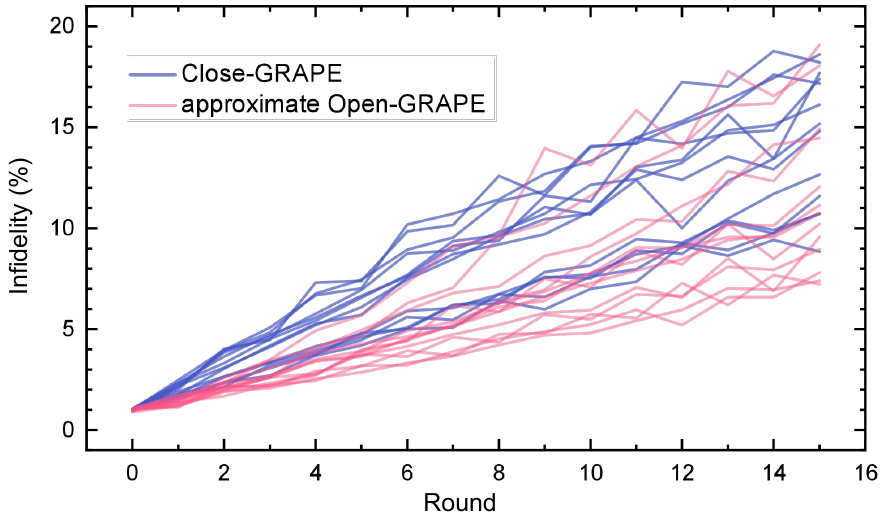}
\par\end{centering}
\caption{\label{fig:FigS3_LinearFitting}The infidelities corresponding to $13$ different pulses versus the number of repetitive $R_y(\pi)$ gate operations in the experiment. The blue lines and the red lines represent the performance of the pulses generated from the Closed-GRAPE and approximate Open-GRAPE algorithms, respectively. For simplicity, the markers of the data points and their standard deviations are not shown.}
\end{figure}

Figure~\ref{fig:FigS3_LinearFitting} shows the infidelity $\widetilde{f}$ of $13$ pulses in the repetitive $R_y(\pi)$ gate experiment. Each line in the figure corresponds to an $R_y(\pi)$ gate operation implemented by different pulses generated by the GRAPE algorithm, while the state initialization and decoding pulses remain the same. According to the figure, most red lines lie below the blue lines, and this indicates that the approximate Open-GRAPE algorithm outperforms the Closed-GRAPE algorithm on average. A comparison of the average performance is also depicted in Fig.~4D in the main text. The infidelities can be obtained by implementing linear fitting on each line, and the average infidelities are $0.89\% \pm 0.21\%$ and $0.72\% \pm 0.28\%$ for the Closed-GRAPE algorithm and the approximate Open-GRAPE algorithm, respectively. The optimal infidelities are $0.58\% \pm 0.02\%$ and  $0.44\% \pm 0.01\%$ for the Closed-GRAPE algorithm and the approximate Open-GRAPE algorithm, respectively.

\end{document}